\documentclass[epj,nopacs]{svjour}
\usepackage{graphics}
\usepackage{graphicx}
\usepackage{tensor}
\usepackage{amsmath}
\usepackage{amssymb}
\usepackage{amsfonts}
\usepackage{url}
\usepackage{xcolor}
\begin{document}
\sloppy

\title{Herglotz-type $f(R,T)$ gravity}

\author{Marek Wazny\inst{1} \and Lehel Csillag\inst{1,2} \and Miguel A. S. Pinto\inst{3,4} \and  Tiberiu Harko\inst{1,5}
}                  
\institute{Department of Physics, Babeș-Bolyai University, Kogălniceanu Street, Cluj-Napoca 400084, Romania \and Faculty of Mathematics and Computer Science, Transilvania University, Iuliu Maniu Street 50, Brașov 500091, Romania  \and Instituto de Astrofísica e Ciências do Espaço, Faculdade de Ciências da Universidade de Lisboa, Edifício C8, Campo Grande, P-1749-016 Lisbon, Portugal \and  Departamento de F\'{i}sica, Faculdade de Ci\^{e}ncias da Universidade de Lisboa, Edifício C8, Campo Grande, P-1749-016 Lisbon, Portugal \and Astronomical Observatory, 19 Cireșilor Street, Cluj-Napoca 400487, Romania}
\date{Received: date / Revised version: date}

\abstract{
 The non-conservation of the energy-momentum tensor in $f(R,T)$ gravity can be interpreted as an effective manifestation of dissipation. Motivated by this, we propose a new formulation of $f(R,T)$ gravity based on the Herglotz variational principle, which extends the usual {Hamilton} variational principle to dissipative systems by allowing the Lagrangian to depend explicitly on the action. The resulting gravitational field equations extend those of $f(R,T)$ gravity by including Herglotz contributions. In the Newtonian limit, these contributions modify the gravitational potential, allowing us to constrain the Herglotz vector through Mercury's perihelion precession and the relativistic light deflection. The Herglotz corrections lead to a scaling law consistent with observations from the Cassini spacecraft. Examining two representative cosmological models, the Herglotz vector effectively reduces to a single function that, under suitable conditions, can play the role of a cosmological constant, providing an alternative mechanism for the Universe's accelerated expansion. Within the Herglotz variational approach, the linear $f(R,T)=R+\alpha T$ model, previously ruled out in the standard formulation due to its fixed deceleration parameter, becomes consistent with observations {for a certain choice of the Herglotz vector}.
}
\maketitle


\section{Introduction}
Conservation laws play a central role in modern theoretical physics, ever since the formulation of the celebrated Noether theorems. These theorems, in fact, arose from discussions among Einstein, Klein, Noether, and Hilbert concerning the relevance of local energy-momentum conservation in general relativity \cite{Brading:2005ina}. Einstein originally derived the gravitational field equations from physically motivated principles -- namely, general covariance, local energy-momentum conservation, and consistency with Newtonian mechanics.  Almost simultaneously with Einstein, Hilbert found a variational formulation for general relativity, in which the corresponding Euler-Lagrange equations coincided with Einstein's field equations. In doing so, both Hilbert and Einstein assumed that geometry and matter were minimally coupled to each other. Then, it was shown that splitting the action into a geometry action and a matter action, alongside the diffeomorphism invariance of the latter, naturally implies the local conservation of the energy-momentum tensor \cite{Landau:1975pou} (see \cite{BarroseSa:2025uxe} for a clarification on this topic). Therefore, non-conservative effects may arise from a violation of coordinate invariance or from the absence of a clear separation between the geometric and matter sectors, allowing for their direct interaction.

The first approach traces back to Einstein, as the simplest way to break local Lorentz invariance is by introducing a fixed background field that defines a preferred frame or direction -- reminiscent of the notion of aether. This can be achieved by introducing a scalar field $T$, interpreted as a cosmic time function, whose timelike gradient establishes a preferred rest frame. Alternatively, one can also describe a preferred frame by considering a timelike unit vector field $u^{\mu}$. Such a field is, in fact, one of the main ingredients of the Einstein-Aether theory \cite{Eling:2004dk}, whose action principle involving a Lagrange multiplier $\lambda$ is formulated as 
\begin{equation}
    S=\kappa \int \mathrm{d}^4x \sqrt{-g} \left(R+\tensor{K}{^\mu^\nu_\alpha_\beta} \tensor{\nabla}{_\mu} \tensor{u}{^\alpha} \tensor{\nabla}{_\nu} \tensor{u}{^\beta} +\lambda( \tensor{u}{^\mu} \tensor{u}{_\mu} -1)\right),
\end{equation}
where $\kappa$ is the effective gravitational constant, $R$ is the usual Ricci scalar and $\tensor{K}{^\mu^\nu_\alpha_\beta}$ is a rank-4 tensor defined as
\begin{equation}
    \tensor{K}{^\mu^\nu_\alpha_\beta}=\tensor{c}{_1} \tensor{g}{^\mu^\nu} \tensor{g}{_\alpha_\beta}+ \tensor{c}{_2} \tensor{\delta}{^\mu_\alpha} \tensor{\delta}{^\nu_\beta}+ \tensor{c}{_3} \tensor{\delta}{^\mu_\beta} \tensor{\delta}{^\nu_\alpha}+ \tensor{c}{_4} \tensor{u}{^\mu} \tensor{u}{^\nu} \tensor{g}{_\alpha_\beta}, 
\end{equation}
with the coefficients $\tensor{c}{_1}, \tensor{c}{_2}, \tensor{c}{_3}, \tensor{c}{_4}$ being real constants. Within Einstein-Aether theory, linearized waves \cite{PhysRevD.70.024003}, PPN parameters \cite{PhysRevD.69.064005}, stars \cite{Eling:2006df,Leon:2019jnu}, black holes \cite{Eling:2006ec,PhysRevD.103.044001} and cosmology \cite{Paliathanasis:2020pax,Paliathanasis:2020axi,Paliathanasis:2019pcl} were explored \cite{Eling:2004dk}. For a generalization in integrable Weyl geometry, see \cite{Paliathanasis:2020plf,Paliathanasis:2021qns} .

The second approach to construct non-conservative theories is through coupling geometry and matter non-minimally. This implies one builds a theory whose action is characterized by having these two sectors no longer separated from each other. As examples of such theories, we mention the $f(R,T)$ \cite{Harko_2011}, $f\left(R,T,\tensor{R}{_{\mu\nu}} \tensor{T}{^{\mu\nu}} \right)$ \cite{PhysRevD.88.044023}, $f(R,\tensor{\mathcal{L}}{_m})$ \cite{Harko:2010mv}, $f(T,\mathcal{T})$ \cite{Harko:2014aja} and $f\left(R,\tensor{T}{_{\mu\nu}} \tensor{T}{^{\mu\nu}}\right)$ \cite{Cipriano:2024jng} gravity theories, where $R$ and $\tensor{R}{_{\mu\nu}}$ are, respectively, the Ricci scalar and Ricci tensor, $T$ is the trace of the energy-momentum tensor of matter $\tensor{T}{_{\mu\nu}}$, defined in terms of the matter Lagrangian $\mathcal{L}_m$, and $\mathcal{T}$ is the torsion scalar. While these do not violate local Lorentz invariance as the Einstein-Aether theory, their non-conservative effects manifest in the emergence of an extra force and in the violation of the equivalence principle due to the general non-conservation of the energy-momentum tensor. Nevertheless, certain conservative models can be constructed \cite{PhysRevD.107.124005,dosSantos2018}. For a comprehensive discussion of energy-momentum conservation in modified gravity theories, see \cite{Koivisto:2005yk}. These theories have been explored in a wide variety of topics, including gravitational waves \cite{Bertolami:2017svl,AZIZI2023101303,PhysRevD.111.024014}, compact objects \cite{dosSantos2018,Carvalho2019,Deb:2018gzt,Pinto:2025loq}, and cosmology \cite{Chakraborty2013,Alvarenga_2013,Tretyakov:2018yph,PhysRevD.98.064045,PhysRevD.102.084051,Farias:2021jdz}. It has also been shown that they possess rich phenomenology, such as the possibility of accounting for irreversible matter creation \cite{Harko_2014,Harko:2015pma,Pinto_2022},  traversable wormholes without exotic matter \cite{Rosa:2022osy,Rosa:2023guo}, and late-time accelerated expansion without the need of dark energy \cite{Bouali_2023,Siggia_2025}.

 Among the broad class of theories incorporating nonminimal couplings between geometry and matter, $f(R,T)$ gravity stands out as one of the most intensively investigated and conceptually appealing formulations. Since this theory is generally non-conservative, as the energy-momentum tensor is not necessarily conserved, it becomes natural to study it from the perspective of Lazo's non-conservative gravity \cite{Lazo_2017,Paiva_2022,Fabris_2018}, which has only been explored in the context of action-dependent Einstein-Hilbert Lagrangians. As such, the non-conservation of the energy-momentum tensor could be interpreted as an effective dissipative process. For a discussion on the role of  dissipation in the scalar field models see \cite{Harko:2023tzo}.

 {More generally, dissipative processes play a fundamental role in nature and are commonly regarded as the physical origin of irreversibility and entropy production in both classical and quantum systems \cite{N1}. In macroscopic systems, irreversibility is typically associated with dissipative effects such as friction and viscosity. However, decay phenomena also represent an important class of irreversible processes. These include radiation emission and nuclear fission-type events, where energy is irreversibly released from the system. In quantum systems, dissipation arises naturally from the interaction between the system and its surrounding environment \cite{N2}. From a thermodynamic perspective, energy decay is usually interpreted as the irreversible transfer of energy from the system to an external environment, often modeled as a termal bath. Nevertheless, it should be emphasized that not all decay mechanisms can be attributed to direct coupling with a thermal reservoir \cite{N3}.}

{The bulk and shear viscous processes are assumed to play a very important role in many astrophysical and cosmological processes, and their role  has been extensively investigated, especially in the early evolution of the Universe \cite{N4}. A rigorous approach to dissipation requires the framework of  causal relativistic thermodynamics, which takes into account the constraints imposed by the finite speed of the thermodynamic interactions. From a cosmological perspective  the interaction between dark matter and dark energy \cite{N5} could provide a possible physical and cosmological mechanism for the presence of the irreversibility and dissipative phenomena in the present day Universe.}

{On the other hand, irreversibility and dissipation is also a characteristic of the gravitational field itself \cite{N7}, especially when coupled to quantum matter \cite{N6}. Gravity is continuously coupled to its matter environment through the dynamical structure of spacetime, and this interaction generically contributes to the growth of entropy in the Universe, thus preventing a strictly reversible evolution. Gravitational wave emission is a general property of gravitational systems, and it essentially represents an energy decay process, leading to dissipative and irreversible behavior. Moreover, the possibility of quantum particle creation in an expanding universe \cite{N8} also indicates the possibility of the irreversible transfer  of energy from the gravitational field to matter. This process is dissipative and is naturally associated with entropy production. Hence, in the presence of gravitational fields, no physical system can be considered perfectly isolated. Irreversibility therefore appears generically in gravitational settings, even though dissipative effects are often neglected within simplified or idealized dynamical descriptions.}

{Taken together, these considerations motivate extending $f(R,T)$ gravity through the Herglotz variational principle, which incorporates dissipative effects.} Introduced in $1930$ by the mathematician Gustav Herglotz in order to describe {frictional} forces\footnote{It should be noted, however, that for example the damped harmonic oscillator can also be formulated without invoking the Herglotz principle, either through standard or non-standard Lagrangians; see \cite{Harko:2023tzo} and references therein. As such, deriving its equations of motion via the Herglotz formalism provides an elegant alternative, motivated by the success of the principle in modeling other dissipative systems.}, this alternative variational principle differs from the Hamiltonian variational principle by allowing the Lagrangian to depend on the action itself \cite{Herg1}. The action dependence of the Lagrangian allows one to describe a broad class of dissipative systems arising in classical mechanics, classical field theory, and quantum mechanics \cite{Lazo_2018,deLeon:2024ztn}. Remarkably, a fully covariant formulation of the Herglotz variational principle -- necessary for its application to field theories -- was developed only in 2018, nearly nine decades after Herglotz first introduced the corresponding variational problem in classical mechanics \cite{Lazo_2018}. Perhaps this covariant formulation was a byproduct of the application of the Herglotz principle in general relativity, since a straightforward application of the original principle led to field equations that were not generally covariant \cite{Lazo_2017}. However, this problem was later remedied \cite{Paiva_2022} by a proper application of the covariant formulation proposed in \cite{Lazo_2018}. For a more rigorous, mathematically detailed derivation of the field equations of nonconservative gravity in fully covariant form, we refer the reader to \cite{ArnauMas}. In this theory, apart from the metric, there is an additional closed one-form $\lambda_{\mu}$ that does not automatically possess dynamics (i.e. is a background field), which in a cosmological setting reduces to a function of time $\phi(t)$ \cite{universe7020038}. The background cosmology and linear perturbative regime of the theory were investigated in \cite{Fabris_2018}, where it was shown that the background dynamics are equivalent to those of bulk viscous cosmology \cite{PhysRevD.53.5483}. Additionally, non-conservative gravity was applied to braneworld models \cite{FabrisBraneWorld}, cosmic string configurations \cite{Braganca2019}, and to the study of spherically symmetric configurations \cite{PhysRevD.99.124031,PhysRevD.103.044018,PhysRevD.110.124056}.

 In this work, we extend $f(R,T)$ gravity by considering it through the Herglotz formalism, allowing for the possibility of {including in a systematic way the basic physical concepts of dissipation and irreversibility in the mathematical structure of the theory, and thus leading to the possibility of constructing new classes of $f(R,T)$ type theories}. 
  {Even though it is rather phenomenological, the Herglotz variational formalism is particularly well adapted for the description of dissipative systems in the presence of irreversibility, since it proposes a general approach for nonconservative Lagrangians, thus allowing the unification in a single framework of large classes of dissipative models and theories.} We further show that this extension can address certain issues present in the literature, such as the cosmological inviability of certain additive models of the form $f(R,T)=g(R)+\lambda h(T)$. Moreover, we also note that $f(R,T)$ gravity formulated through the Herglotz approach naturally generalizes Lazo's non-conservative gravity, recovering it in the particular case $f(R,T)=R$.

 The paper is structured as follows. In Section \ref{section2}, we review the Herglotz variational principle and illustrate how it can describe dissipative systems in classical mechanics, classical field theory, and quantum mechanics. We then present Lazo’s formulation of general relativity within the Herglotz variational framework. In Section \ref{section3}, we apply the Herglotz variational principle to  $f(R,T)$ gravity, explicitly separating the additional contributions of the Herglotz field in the field equations. This makes transparent how both standard $f(R,T)$ gravity and Lazo's non-conservative gravity are recovered as special cases. We compute the covariant divergence of the energy-momentum tensor and show that, under specific conditions on the Herglotz field, it remains covariantly conserved. The equations of motion for massive particles are derived, and in the weak-field approximation, we obtain a generalized Poisson equation along with exact solutions and corrections to the Newtonian potential. Finally, by considering the perihelion precession of Mercury and light bending, we establish observational bounds on the Herglotz field. In Section \ref{section4}, we study two simple cosmological models in Herglotz-type $f(R,T)$ gravity, considering two functional forms $f(R,T)=R+\alpha T$ and $f(R,T)=R+\alpha T^{-1}$. Because the background Herglotz field is non-dynamical, an additional assumption is required to close the under-determined system. Here, we adopt a linear effective equation of state, $p_{eff}=w \rho_{eff}$. We obtain both analytical and numerical solutions, and show that the linear model $f(R,T)=R+\alpha T$, which is not viable in the standard Lagrangian framework due to its constant deceleration parameter, can become a realistic candidate for describing observational data from cosmic chronometers for an appropriate range of model parameters. In Section \ref{section6}, we conclude the paper with a discussion of our results and provide an outlook on potential directions for future research in this field.

\section{Herglotz variational principle for dissipative dynamics} \label{section2}

In this section, we briefly review the Herglotz variational principle in classical mechanics, in classical field theories, and then in the specific case of general relativity. We discuss both the theoretical aspects, as well as the applications of the theory for the description of the dissipative dynamics of physical systems.

\subsection{The Herglotz variational problem in classical physics}   

The principle of stationary action has become the foundation for many branches of modern physics, including gravitation. In the most standard formulation, this principle is given by the  variational problem 
\begin{align}\label{action}
\begin{split}
    \delta S  &= 0 ,\\
    S[q(t)] &= \int_a^b \limits L\left(q(t), \dot{q}(t), t\right) dt ,
\end{split}
\end{align}
where $S$ is the action, $L$ is the Lagrangian, and $q(t)$ is a generalized coordinate defined over the time interval $[a,b]\in \mathbb{R}$. The solution to this problem determines the function $q(t)$, which makes the action $S$ stationary, via the Euler–Lagrange equations 
\begin{equation}\label{ELeqs}
    \frac{\partial L}{\partial q} -\frac{d}{dt}\frac{\partial L}{\partial \dot{q}} = 0 .
\end{equation}
It is worth mentioning that the second line in Eq. (\ref{action}) can be obtained from the differential equation
\begin{equation}\label{DEaction}
    \dot{S}(t) = L(q(t), \dot{q}(t), t ),
\end{equation}
after integration with boundary conditions
\begin{equation}
    S(a) = \tensor{S}{_a}, \quad q(a) = \tensor{q}{_a}, \quad q(b)=\tensor{q}{_b} ,
\end{equation}
with all these constants being real.

However, it is known that this method has difficulties accounting for dissipative systems, since it cannot produce equations of motion with first-order time derivatives \cite{Bauer1931-ek}. Hence, in order to describe these systems, Herglotz formulated an alternative variational principle that deviates from the standard one by letting the Lagrangian depend on the action itself, as well as the usual dynamical variables \cite{Herg1}.

In this case, it becomes clear that 
\begin{equation}\label{HerglotzDEaction}
    \dot{S} = L(q(t), \dot{q}(t), S(t), t) ,
\end{equation}
can no longer be integrated as in Eq. (\ref{DEaction}) to obtain Hamilton's principle. If instead, we consider some variation $\delta q$ directly from Eq. (\ref{HerglotzDEaction}), we arrive at
\begin{align}\label{zetaDE}
    \delta\dot{S} = \frac{\partial L}{\partial q}\delta q+\frac{\partial L}{\partial\dot{q}}\delta\dot{q} +\frac{\partial L}{\partial S}\delta S,
\end{align}
which can be solved as a differential equation \cite{Herg2}. However, defining the scalar
\begin{equation}
    \psi \equiv \int \frac{\partial L}{\partial S} \,dt ,
\end{equation}
we see that Eq. (\ref{zetaDE}) is equivalent to 
\begin{equation}
    \frac{d}{dt}(e^{-\psi}\delta S) = e^{-\psi}\left(\frac{\partial L}{\partial q}\delta q+\frac{\partial L}{\partial \dot{q}}\delta\dot{q}\right) ,
\end{equation}
which can now be formally integrated. Under the assumption of a stationary action, the left-hand side vanishes and the right-hand side leads to the so-called generalized Euler-Lagrange equations \cite{Herg2}
\begin{equation}\label{HerglotzELeqs}
    \frac{\partial L}{\partial q}-\frac{d}{dt}\frac{\partial L}{\partial \dot{q}}+\gamma\frac{\partial L}{\partial \dot{q}} = 0 ,
\end{equation}
which reduce to Eq. (\ref{ELeqs}) in the case where $\gamma \equiv \partial L/\partial S = 0$.

To observe how non-conservative systems can be derived from this formalism, consider the Lagrangian 
\begin{equation}\label{particleL}
    L = \frac{1}{2}m\dot{x}^2-U(x) -\frac{\nu}{m}S,
\end{equation}
with generalized coordinate $x$. This describes a particle of mass $m$ in a potential $U(x)$, subject to a viscous force that is proportional to a drag coefficient $\nu$. Using Eq. (\ref{HerglotzELeqs}), the equation of motion for this particle becomes
\begin{equation}\label{particleLeom}
    m\ddot{x} +\nu\dot{x} = -\frac{dU}{dx} .
\end{equation}
This is exactly a dissipative system due to the drag term which, depending on external force $F=-dU/dx$, could model a damped harmonic oscillator, for example. Furthermore, we see that the Lagrangian in Eq. (\ref{particleL}) admits equations of motion proportional to first order derivatives, yet still retains a physical Hamiltonian in the sense that it is the sum of all the energies, a feature unique to the Herglotz variational principle \cite{Fred1996,Lazo2014}. For example, the corresponding Hamiltonian for Eq. (\ref{particleL}) becomes
\begin{equation}
    \mathcal{H} = \dot{q}p-L = \frac{1}{2}m\dot{x}^2 +U(x) +\frac{\nu}{m}S ,
\end{equation}
where the canonical variables $q=x$ and $p = \partial L/\partial \dot{x} = m\dot{x}$ take on the usual definitions .

\subsection{The Herglotz variational problem in classical field theories}   

The simple point particle description of the Herglotz variational principle can also be applied to fields. To do so, one promotes all generalized coordinates to fields $ \phi\left(\tensor{x}{^\mu} \right)$, and defines an action density $\tensor{s}{^\mu}$ by
\begin{equation}\label{actiondensity}
    S = \int \tensor{n}{_\mu} \tensor{s}{^\mu}\, \tensor{d}{^{n-1}}x = \int \tensor{\partial}{_\mu} \tensor{s}{^\mu} \,\tensor{d}{^n}x,
\end{equation}
where the divergence theorem has been used. Then, the extension of Eq. (\ref{HerglotzDEaction}) becomes
\begin{equation}
    \tensor{\partial}{_\mu} \tensor{s}{^\mu} = \mathcal{L}\left(\phi\left(\tensor{x}{^\mu}\right), \tensor{\partial}{_\nu} \phi\left(\tensor{x}{^\mu}\right), \tensor{s}{^\mu}, \tensor{x}{^\mu}\right),
\end{equation}
with the appropriate boundary conditions on $\phi$. Now, defining ${\varphi}$, the action density, as
\begin{equation}
    {\varphi} \equiv \int \frac{\partial \mathcal{L}}{\partial \tensor{s}{^\mu}} d \tensor{x}{^\mu} ,
\end{equation}
one arrives at
\begin{align}
    \tensor{\partial}{_\mu} \left(e^{-{\varphi}}\delta \tensor{s}{^\mu} \right) = e^{-{\varphi}}\left(\frac{\partial \mathcal{L}}{\partial \phi}\delta \phi + \frac{\partial \mathcal{L}}{\partial\left( \tensor{\partial}{_\mu} \phi \right)}\delta\left( \tensor{\partial}{_\mu} \phi \right)\right) .
\end{align}
The left hand side vanishes by virtue of $\delta \tensor{s}{^\mu} = 0$, since for stationary actions Eq. (\ref{actiondensity}) gives
\begin{equation}
    \delta S = \int \tensor{\partial}{_\mu} \delta \tensor{s}{^\mu} \,\tensor{d}{^n}x = 0 .
\end{equation}
Then, the right hand side gives the generalized Euler-Lagrange equations for fields \cite{Lazo_2018}
\begin{equation}\label{HerglotzFieldELeq}
    \frac{\partial \mathcal{L}}{\partial \phi} - \tensor{\partial}{_\mu}\frac{\partial \mathcal{L}}{\partial\left( \tensor{\partial}{_\mu} \phi \right)}+ \tensor{\gamma}{_\mu} \frac{\partial \mathcal{L}}{\partial\left( \tensor{\partial}{_\mu} \phi \right)} = 0 ,
\end{equation}
where $\tensor{\gamma}{_\mu} \equiv \partial \mathcal{L}/\partial \tensor{s}{^\mu}$. 

As a simple example, consider the case of a non-conservative electromagnetic theory with Lagrangian density
\begin{equation}
    \mathcal{L} = -\frac{1}{4}\tensor{F}{^{\mu\nu}} \tensor{F}{_{\mu\nu}} +\tensor{J}{^\mu} \tensor{A}{_\mu} -\tensor{\gamma}{_\mu} \tensor{s}{^\mu} ,
\end{equation}
where $\tensor{F}{_{\mu\nu}}= \tensor{\partial}{_\mu} \tensor{A}{_\nu} -\tensor{\partial}{_\nu} \tensor{A}{_\mu}$ is the field strength tensor, $\tensor{A}{_\mu}$ is the four-potential, $\tensor{J}{_\mu}$ is the four-current density, and $\tensor{\gamma}{_\mu}$ is a constant four vector. From this Lagrangian, Eq. (\ref{HerglotzFieldELeq}) gives the equation of motion \footnote{We use the metric signature convention $(-,+,+,+)$ exclusively throughout this work.}
\begin{equation}
    \left(\tensor{\partial}{_\mu} +\tensor{\gamma}{_\mu}  \right)\tensor{F}{^{\mu\nu}}= -\tensor{J}{^\mu} .
\end{equation} 
This dissipative equation can be used to analyze the behavior of an electron in a non-ideal conductor, as well as modifications to cyclotron radiation \cite{doi:10.1142/S0219887822501560}.

As an additional example, one can construct a dissipative Schrödinger-like system with mass $m$ in a potential $U(\vec{r})$. If we separate the covariant notation as $\tensor{x}{_\mu} = \left(t, \vec{r}\right)$ and $\tensor{\partial}{_\mu} =\left(\tensor{\partial}{_t},\vec{\nabla} \right)$, then the Lagrangian density with complex function $\Psi\left(t,\vec{r} \right)$ becomes 
\begin{eqnarray}\label{schrlag}
    \mathcal{L} &=& -\frac{1}{2m}\vec{\nabla}\Psi^*\vec\nabla\Psi -U \Psi^*\Psi +\frac{i}{2}\left(\Psi^*\tensor{\partial}{_t}\Psi - \Psi \tensor{\partial}{_t}\Psi^* \right) \nonumber \\
    &&-\tensor{\gamma}{_\mu} \tensor{s}{^\mu} ,
\end{eqnarray}
where again $\tensor{\gamma}{_\mu} =\left(\tensor{\gamma}{_0}, \vec{\gamma} \right)$ is a constant four vector, and we have assumed $\hbar =1$. Now, applying Eq. (\ref{HerglotzFieldELeq}) to Eq. (\ref{schrlag}) on the complex field $\Psi^*(t,\vec{r})$, we arrive at the dissipative Schrödinger equation
\begin{equation}\label{schreom}
    i \tensor{\partial}{_t} \Psi = \left(-\frac{1}{2m}\nabla^2 +U-\frac{i}{2} \tensor{\gamma}{_0} -\frac{1}{2m}\vec{\gamma}\cdot\vec\nabla\right)\Psi .
\end{equation}

\subsection{Herglotz variational principle in General Relativity}

The Herglotz principle within the framework of general relativity has been explored in Refs. \cite{Lazo_2017} and \cite{Paiva_2022}, which we will follow closely in this section. The results discussed for fields generalize covariantly in the same way as the usual Lagrangian formalism. Thus, the Herglotz variational problem in GR is given by 
\begin{align}\label{GRsys}
    \begin{split}
        \tensor{\nabla}{_\mu}  \tensor{s}{^\mu} &= \mathcal{L}\left(\tensor{g}{_{\alpha\beta}} \left(\tensor{x}{^\mu}\right), \tensor{\partial}{_\mu} \tensor{g}{_{\alpha\beta}}\left( \tensor{x}{^\mu} \right), \tensor{s}{^\mu}, \tensor{x}{^\mu}\right) ,\\
        S(\Omega) &= \int_\Omega \limits \tensor{d}{^{n-1}}x\sqrt{h} \, \tensor{n}{_\mu} \tensor{s}{^\mu} = \int_\mathcal{V}\limits  \tensor{d}{^n}x \sqrt{-g} ~ \tensor{\nabla}{_\mu} \tensor{s}{^\mu},
    \end{split}
\end{align}
defined on a $n$-dimensional smooth manifold $\mathcal{M}$, with metric $\tensor{g}{_{\mu\nu}}$. Let $\mathcal{V}$ be a subset of $\mathcal{M}$ with boundary $\Omega$, having  unit normal vector $\tensor{n}{_\mu}$. The determinants of the metrics $\sqrt{h}$ and $\sqrt{-g}$ have the usual definition. 

{To implement the covariant Herglotz formalism, we introduce a one-form
\begin{equation}
    \lambda\equiv\lambda_{\mu} dx^{\mu},
\end{equation}
whose components $\lambda_{\mu}$ depend solely on the spacetime coordinates, and plays the role of a generalized dissipative coefficient. By  construction $\lambda_{\mu}$ is assumed to be a closed one-form, 
\begin{equation}
    \nabla_{\mu} \lambda_{\nu} - \nabla_{\nu} \lambda_{\mu}=0,
\end{equation}
so that locally it can be written as
\begin{equation}
    \lambda_{\mu}=\partial_{\mu} \chi,
\end{equation}
for some scalar field $\chi$. In this framework, $\lambda_{\mu}$ does not possess independent dynamics, hence it must be prescribed.
}

{
We now consider the Lagrangian density
}
\begin{equation}
    \mathcal{L} = R+F\tensor{\mathcal{L}}{_m} + \tensor{\lambda}{_\mu} \tensor{s}{^\mu},
\end{equation}
where $R$ is the Ricci scalar, $\tensor{\mathcal{L}}{_m}$ is the matter Lagrangian, and $F=F(x)$ is a coupling constant potentially depending on coordinates. Since $\mathcal{L}$ contains second order derivatives of the metric we impose boundary conditions $\tensor{g}{_{\mu\nu}}$ and $\tensor{\partial}{_\rho} \tensor{g}{_{\mu\nu}}$ fixed on $\Omega$. This leads to the variations
\begin{align}\label{GR}
\begin{split}
    \delta S(\Omega) &= \int_\Omega \limits \tensor{d}{^{n-1}}x\, \tensor{n}{_\mu}\delta\left(\sqrt{h} \tensor{s}{^\mu}\right)=0 ,\\
    \tensor{\partial}{_\mu}\left(e^{-{\varphi}}\delta(\sqrt{-g} \tensor{s}{^\mu})\right)&= e^{-{\varphi}}\delta\left(\sqrt{-g}R+\sqrt{-g}F \tensor{\mathcal{L}}{_m}\right) ,
\end{split}
\end{align}
with 
\begin{equation}
    {\varphi =\int \frac{\partial \mathcal{L}}{\partial s^{\mu}} dx^{\mu}}= \int \tensor{dx}{^\mu} \tensor{ \lambda}{_\mu} .
\end{equation}

By fixing $\Omega$ and $\sqrt{h}$, the first equation in (\ref{GR}) ensures that $\delta \tensor{s}{^\mu}=0$. Focusing on the second equation, the left hand side reads after integration 
\begin{align}
       &\int_\mathcal{V}\limits \tensor{d}{^n} x\,  \tensor{\partial}{_\mu} \left(e^{-{\varphi}} \delta(\sqrt{-g}\tensor{s}{^\mu})\right)  \nonumber\\
       &=  \int_\Omega \limits \tensor{d}{^{n-1}}x\sqrt{h}\, e^{-{\varphi}} \tensor{n}{_\mu}  \left(\delta \tensor{s}{^\mu} -\frac{1}{2}\tensor{s}{^\mu} \tensor{g}{_{\rho \nu}} \delta \tensor{g}{^{\rho\nu}}\right)  \nonumber\\
       &=0 ,
\end{align}
since the metric and the action density field are fixed on $\Omega$. Taking this result, the right hand side of the equation gives
\begin{align}\label{GRvar}
   0&= \int_{\mathcal{V}} \limits  \tensor{d}{^n} x \, e^{-{\varphi}} \delta \left( \sqrt{-g} R + \sqrt{-g} F \tensor{\mathcal{L}}{_{m}} \right) \nonumber\\
   &=\int_{\mathcal{V}}\limits \tensor{d}{^n} x \sqrt{-g} e^{- {\varphi} }  \left( \tensor{G}{_{\mu \nu}} + \tensor{K}{_{\mu \nu}} - \frac{F}{2} \tensor{T}{_{\mu\nu}}\right)\delta \tensor{g}{^{\mu \nu}} ,
\end{align}
where we have defined the tensors
\begin{align}
    \tensor{G}{_{\mu\nu}} &\equiv \tensor{R}{_{\mu\nu}}-\frac{1}{2}\tensor{g}{_{\mu\nu}}R ,\\
    \tensor{T}{_{\mu\nu}} &\equiv -\frac{2}{\sqrt{-g}}\frac{\delta(\sqrt{-g} \tensor{\mathcal{L}}{_m})}{\delta \tensor{g}{^{\mu\nu}}} ,\\
    \tensor{K}{_{\mu\nu}} &\equiv \frac{1}{2}\left( \tensor{\nabla}{_\mu} \tensor{\lambda}{_\nu} + \tensor{\nabla}{_\nu} \tensor{\lambda}{_\mu}\right) -\tensor{\lambda}{_\mu} \tensor{\lambda}{_\nu} - \tensor{g}{_{\mu\nu}}\left( \tensor{\nabla}{_\rho} \tensor{\lambda}{^\rho}-\tensor{\lambda}{_\rho} \tensor{\lambda}{^\rho}\right) .
\end{align}
From Eq. (\ref{GRvar}), we find the field equations using the Herglotz variational principle
\begin{equation}
    \tensor{G}{_{\mu \nu}} + \tensor{K}{_{\mu \nu}} = \frac{F}{2} \tensor{T}{_{\mu\nu}} ,
\end{equation}
which reduce to the Einstein field equations when $\tensor{\lambda}{_\mu}=0$, $F=16\pi$.

\section{Generalized  $f(R,T)$ gravity via the Herglotz variational principle} \label{section3}

In this section, we introduce a new formulation of $f(R,T)$ gravity based on the Herglotz variational principle. We derive the covariant divergence of the matter energy-momentum tensor and obtain the corresponding energy balance, as well as the equations of motion for massive particles, assuming a perfect fluid matter distribution.
\subsection{Gravitational action and field equations}

We consider the same system as (\ref{GRsys}), but with $\mathcal{L} = f(R,T)+ \tensor{\lambda}{_\mu} \tensor{s}{^\mu} + F \tensor{\mathcal{L}}{_m}$. {Given $\mathcal{L}_{m}$, we define
\begin{equation}
    T_{\mu \nu}\equiv-\frac{2}{\sqrt{-g}}\frac{\delta \left( \sqrt{-g} \mathcal{L}_{m} \right)}{\delta g^{\mu \nu}} ,
\end{equation}
with $T\equiv T_{\mu \nu} g^{\mu \nu}$ its trace. If $\mathcal{L}_{m}$ depends only on the metric, and not on its derivatives, we obtain
\begin{equation}
    T_{\mu \nu}=g_{\mu \nu} \mathcal{L}_{m}-2\frac{\partial \mathcal{L}_{m}}{\partial g^{\mu \nu}}.
\end{equation}\
}
Consequently, when taking variations we need only consider 
\begin{align} \label{eq:variations_F_R_T}
    0&= \int_\mathcal{V} \limits \tensor{d}{^n}x\,e^{-{\varphi}}\delta\left(\sqrt{-g}f(R,T)+\sqrt{-g} \tensor{\mathcal{L}}{_m}\right) \nonumber\\
     &= \int_\mathcal{V} \limits \tensor{d}{^n} x\sqrt{-g} e^{-{\varphi}}\bigg[f_R(R,T)\tensor{K}{_{\mu\nu}}+ \tensor{\widetilde{K}}{_{\mu\nu}}+ f_R(R,T) \tensor{R}{_{\mu\nu}} \nonumber\\
     &+\left(\tensor{T}{_{\mu\nu}}+\tensor{\Theta}{_{\mu\nu}}\right)\tensor{f}{_T}(R,T)-\frac{1}{2}\tensor{g}{_{\mu\nu}} f(R,T) -\frac{F}{2}\tensor{T}{_{\mu\nu}}\bigg]\delta \tensor{g}{^{\mu\nu}},
\end{align}
where we have denoted subscripts as partial derivatives and defined 
\begin{align}
     \Theta_{\mu \nu}&\equiv g^{\alpha \beta} \frac{ \delta T_{\alpha \beta}}{\delta g^{\mu \nu}}, \; \; T=T^{\mu\nu} g_{\mu \nu}, \\
     \tensor{\widetilde{K}}{_{\mu\nu}} &\equiv \left(\tensor{g}{_{\mu \nu}} \Box  - \tensor{\nabla}{_\mu} \tensor{\nabla}{_\nu} +\tensor{\lambda}{_\mu} \tensor{\partial}{_\nu}  +\tensor{\lambda}{_\nu} \tensor{\partial}{_\mu}  - 2\tensor{g}{_{\mu\nu}} \tensor{\lambda}{^\rho} \tensor{\partial}{_\rho} \right) \tensor{f}{_R} ,
\end{align}
with $\Box \equiv \tensor{\nabla}{_\mu} \tensor{\nabla}{^\mu}$. {We note that if $\mathcal{L}_{m}$ depends only on the metric, and not on its derivatives, $\Theta_{\mu \nu}$ can be computed as \cite{Harko_2011}
\begin{equation}\label{eq:Theta_munu}
    \Theta_{\mu \nu}=-2 T_{\mu \nu}+ g_{\mu \nu} \mathcal{L}_{m} - 2g^{\alpha \beta} \frac{\partial^2 \mathcal{L}_{m}}{\partial g^{\mu \nu} \partial g^{\alpha \beta}}.
\end{equation}
}Dropping the dependence on $R$ and $T$, the full variational principle gives 
\begin{align}
    &\tensor{f}{_{R}} \tensor{R}{_{\mu \nu}} -\frac{1}{2} \tensor{g}{_{\mu \nu}} f + \tensor{f}{_R} \tensor{K}{_{\mu \nu}} + \tensor{\widetilde{K}}{_{\mu\nu}} \nonumber\\
    &= \frac{F}{2}  \tensor{T}{_{\mu \nu}} - \tensor{T}{_{\mu \nu}} \tensor{f}{_{T}} - \tensor{\Theta}{_{\mu \nu}} \tensor{f}{_{T}}.
\end{align}

In addition to how $f(R,T)$ affects the usual Herglotz derivation in general relativity, we can more clearly see the contribution of the Herglotz field to $f(R,T)$ gravity, 
\begin{align}\label{Herglotzeom}
    &\tensor{f}{_{R}} \tensor{R}{_{\mu \nu}} - \frac{1}{2} f \tensor{g}{_{\mu\nu}} + \left(  \tensor{g}{_{\mu \nu}} \Box - \tensor{\nabla}{_{\mu}} \tensor{\nabla}{_{\nu}} \right) \tensor{f}{_{R}}+\tensor{H}{_{\mu \nu}} \nonumber\\
    &=\frac{F}{2} \tensor{T}{_{\mu \nu}} - \tensor{T}{_{\mu \nu}}  \tensor{f}{_{T}} - \tensor{\Theta}{_{\mu \nu}} \tensor{f}{_{T}},
\end{align}
where we have introduced the Herglotz tensor
\begin{equation}
    \tensor{H}{_{\mu\nu}}=\tensor{f}{_R} \tensor{K}{_{\mu \nu}}+ \tensor{\lambda}{_\mu}  \tensor{\partial}{_\nu} \tensor{f}{_R} +\tensor{\lambda}{_\nu} \tensor{\partial}{_\mu} \tensor{f}{_R} - 2\tensor{g}{_{\mu\nu}} \tensor{\lambda}{_\rho} \tensor{\partial}{^\rho} \tensor{f}{_R} .
\end{equation}
The following remarks are in order:
\begin{enumerate}
    \item For a vanishing Herglotz field. $\tensor{\lambda}{_\mu}=0$, and a constant $F(x)=16\pi$, the field equations reduce to those of standard $f(R,T)$ gravity \cite{Harko_2011}.
    \item For a non-vanishing Herglotz field, $\tensor{\lambda}{_\mu} \neq 0$, and $f(R,T)=R$, the field equations reproduce the non-conservative gravity of Lazo et. al. \cite{Paiva_2022}.
    \item For $\lambda_{\mu}=0,f(R,T)=R$ and $F(x)=16\pi$, the field equations reduce to the Einstein equation of general relativity.
\end{enumerate}

By taking the trace of Eq.~(\ref{Herglotzeom}) we obtain 
\begin{equation}\label{trace}
-\frac{1}{2}f=-\frac{1}{4}\tensor{f}{_R}R-\frac{3}{4}\Box \tensor{f}{_R}-\frac{1}{4}H+\frac{F}{8}T-\frac{1}{4}\tensor{f}{_T}\left(T+\Theta\right),
\end{equation}
where $H=\tensor{H}{_\mu^\mu}$, $\Theta =\tensor{\Theta}{ _\mu ^\mu}$, and $T=\tensor{T}{_\mu^\mu}$. Eq.~(\ref{trace})  allows us to reformulate the field equation (\ref{Herglotzeom}) of Herglotz-type $f(R,T)$ gravity in the traceless form
\begin{align} \label{tracelessform}
&\tensor{f}{_R}\left(\tensor{R}{_{\mu \nu}}-\frac{1}{4}R \tensor{g}{_{\mu \nu}}\right)+\left(\frac{1}{4}\tensor{g}{_{\mu \nu}}\Box -\tensor{\nabla}{_\mu} \tensor{\nabla}{_\nu}\right)\tensor{f}{_R} \nonumber\\
&+\tensor{H}{_{\mu \nu}}-\frac{1}{4}H \tensor{g}{_{\mu \nu}}=\frac{F}{2}\left( \tensor{T}{_{\mu \nu}}-\frac{1}{4}T \tensor{g}{_{\mu \nu}}\right)\nonumber\\
&-\left[\left(\tensor{T}{_{\mu \nu}}+\tensor{\Theta}{_{\mu \nu}}\right)-\frac{1}{4}\left(T+\Theta\right) \tensor{g}{_{\mu \nu}}\right] \tensor{f}{_T}.
\end{align}

\subsection{Conservation of the energy-momentum tensor}

In general, within $f(R,T)$ gravity, the energy-momentum tensor is not conserved. This non-conservation has been interpreted as either particle creation or annihilation processes \cite{Singh2015,Singh2016,Harko_2014,Asadiyan_2019}, or more formally, as particle production arising from the thermodynamics of open systems \cite{Pinto_2022,Bouali_2023}. Alternatively, one may impose specific conditions on the function $f(R,T)$ to enforce the conservation of the energy-momentum tensor \cite{Carvalho2019,dosSantos2018,Alvarenga_2013,Chakraborty2013}. In the following, we determine conditions on the Herglotz field that guarantee energy-momentum conservation.

To find the energy-balance equation, we take the divergence on both sides of (\ref{Herglotzeom}). The left hand side gives 
\begin{align}
   & \tensor{\nabla}{^\mu} \left[\tensor{f}{_{R}} \tensor{R}{_{\mu \nu}} - \frac{1}{2} f \tensor{g}{_{\mu\nu}} + \left(  \tensor{g}{_{\mu \nu}} \Box - \tensor{\nabla}{_{\mu}} \tensor{\nabla}{_{\nu}} \right) \tensor{f}{_{R}}+\tensor{H}{_{\mu \nu}}\right] \nonumber\\
    &=  \tensor{\nabla}{^\mu} \tensor{H}{_{\mu\nu}} -\frac{1}{2}\tensor{f}{_T} \tensor{\nabla}{_\nu} T ,
\end{align}
where we used the identities $\left[ \tensor{\nabla}{_\mu}, \tensor{\nabla}{_\nu} \right] \tensor{\nabla}{^\mu} \tensor{f}{_R}+\tensor{R}{_{\mu\nu}} \tensor{\nabla}{^\mu} \tensor{f}{_R} = 0$ and $  \tensor{\nabla}{^\mu} \tensor{G}{_{\mu\nu}}=0$. The right hand side reads 
\begin{align}
&\tensor{\nabla}{^\mu} \left[\frac{F}{2} \tensor{T}{_{\mu \nu}} - \tensor{T}{_{\mu \nu}}  \tensor{f}{_{T}} - \tensor{\Theta}{_{\mu \nu}} \tensor{f}{_{T}}\right] \nonumber\\
    &= \left(\frac{F}{2}+\tensor{f}{_T}\right) \tensor{\nabla}{^\mu} \tensor{T}{_{\mu\nu}}+\frac{1}{2}\tensor{T}{_{\mu\nu}} \tensor{\nabla}{^\mu} F -\left(\tensor{T}{_{\mu\nu}}+ \tensor{\Theta}{_{\mu\nu}}\right) \tensor{\nabla}{^\mu}  \tensor{f}{_T} \nonumber\\
    &-\tensor{f}{_T}\left( \tensor{\nabla}{_\nu} \tensor{\mathcal{L}}{_m} -2\tensor{g}{^{\alpha\beta}} \tensor{\nabla}{^\mu} \frac{\partial ^2 \tensor{\mathcal{L}}{_m}}{\partial \tensor{g}{^{\mu \nu}} \partial  \tensor{g}{^{\alpha \beta}}} \right) ,
\end{align}
where we used {the identity \eqref{eq:Theta_munu}.}

Putting these results together gives the divergence of the energy-momentum tensor
\begin{eqnarray}\label{EMdiv}
    \tensor{\nabla}{^\mu} \tensor{T}{_{\mu\nu}} &=& \frac{2}{F+2 \tensor{f}{_T}}\bigg[ \tensor{\nabla}{^\mu} \tensor{H}{_{\mu\nu}}-\frac{1}{2}\tensor{T}{_{\mu\nu}} \tensor{\nabla}{^\mu} F \nonumber\\
   && +\frac{1}{2}\tensor{f}{_T} \tensor{\nabla}{_\nu}(2 \tensor{\mathcal{L}}{_m}-T) 
    +\left(\tensor{T}{_{\mu\nu}}+ \tensor{\Theta}{_{\mu\nu}}\right) \tensor{\nabla}{^\mu} \tensor{f}{_T} \nonumber\\
    &&-2\tensor{g}{^{\alpha\beta}} \tensor{f}{_T} \tensor{\nabla}{^\mu}\frac{\partial ^2 \tensor{\mathcal{L}}{_m}}{\partial\tensor{g}{^{\mu \nu}} \partial  \tensor{g}{^{\alpha \beta}}} \bigg] \equiv \tensor{J}{_\nu}.
\end{eqnarray}

Similar to the regular $f(R,T)$ gravity case, Eq. (\ref{EMdiv}) shows that the divergence does not vanish in general. However, it is possible to impose a condition on the Herglotz field such that the energy-momentum tensor becomes conserved. In this case, the Herglotz field is subject to the constraint
\begin{eqnarray}\label{Herglotzdiv}
    \tensor{\nabla}{^\mu} \tensor{H}{_{\mu\nu}}&=&\frac{1}{2}\tensor{T}{_{\mu\nu}} \tensor{\nabla}{^\mu} F-\frac{1}{2}\tensor{f}{_T} \tensor{\nabla}{_\nu}(2 \tensor{\mathcal{L}}{_m}-T) \nonumber\\
    &&-(\tensor{T}{_{\mu\nu}}+ \tensor{\Theta}{_{\mu\nu}}) \tensor{\nabla}{^\mu} \tensor{f}{_T} +2 \tensor{g}{^{\alpha\beta}} \tensor{f}{_T} \tensor{\nabla}{^\mu}\frac{\partial ^2 \tensor{\mathcal{L}}{_m}}{\partial \tensor{g}{^{\mu \nu}} \partial  \tensor{g}{^{\alpha \beta}}}.\nonumber\\
\end{eqnarray}

\subsection{Energy balance and equation of motion}

In the following, we assume that the matter content of the gravitating system consists of a perfect fluid, with  {$\mathcal{L}_{m}=p$} and energy-momentum tensor given by
\begin{equation}\label{perfT}
\tensor{T}{_{\mu\nu}}=(\rho+p)\tensor{u}{_\mu} \tensor{u}{_\nu} +p \tensor{g}{_{\mu\nu}},
\end{equation}
where $\rho$ is the total baryonic matter energy density, and $p$ is the thermodynamic pressure, respectively. The four-velocity $\tensor{u}{^\mu}=\tensor{dx}{^\mu}/ds$ of the massive object is introduced as the tangent vector of the worldline of the particle, parameterized  by the arc length $s$. Hence the four-velocity satisfies the normalization condition $\tensor{u}{_\mu} \tensor{u}{^\mu}=-1$.
We take now the covariant derivative of Eq.~(\ref{perfT}), thus obtaining
\begin{eqnarray}\label{perfdivT}
 \tensor{\nabla}{^\mu} \tensor{T}{_{\mu\nu}}=\tensor{J}{_\nu}&=&\left(\tensor{\nabla}{^\mu} p+\tensor{\nabla}{^\mu} \rho \right) \tensor{u}{_\mu} \tensor{u}{_\nu}+(p+\rho)\times \nonumber\\
 &&\left( \tensor{u}{_\nu} \tensor{\nabla}{^\mu} \tensor{u}{_\mu} +\tensor{u}{_\mu} \tensor{\nabla}{^\mu} \tensor{u}{_\nu} \right) +\tensor{\nabla}{_\nu} p.
\end{eqnarray}
After multiplying both sides of the above relation with $\tensor{u}{^\nu}$, we obtain the energy balance equation in Herglotz-type $f(R,T)$ gravity as
\begin{equation}
-\tensor{u}{_\mu} \tensor{\nabla}{^\mu} \rho-(p+\rho)\tensor{\nabla}{^\mu} \tensor{u}{_\mu} \equiv -\dot{\rho}-3H(p+\rho)=\tensor{u}{^\nu} \tensor{J}{_\nu},
 \end{equation}
where we have taken into account the relation $\tensor{u}{^{\mu}} \tensor{\nabla}{_{\nu}} \tensor{u}{_{\mu}}=0$, and we have defined the covariant Hubble function $H$ as $3H\equiv \tensor{\nabla}{^\mu} \tensor{u}{_\mu}$. The dot denotes $\tensor{u}{_\mu} \tensor{\nabla}{^\mu}=d/ds$.

Hence the energy source $\mathcal{S}$ in the gravitating system is given by
\begin{equation}
    \dot{\rho}+3H(p+\rho)=-\tensor{u}{^\nu} \tensor{J}{_\nu}\equiv \mathcal{S}.
\end{equation}

Multiplying both sides of Eq.~(\ref{perfdivT}) with the projection operator 
\begin{equation}
\tensor{h}{^{\nu\rho}}\equiv \tensor{g}{^{\nu\rho}}+\tensor{u}{^\nu} \tensor{u}{^\rho},
\end{equation}
we find the momentum balance equation (or the equation of motion)
\begin{equation}\label{eqmot}
    \tensor{u}{^\mu} \tensor{\nabla}{_\mu} \tensor{u}{^\rho}=\frac{d^2 \tensor{x}{^{\rho}}}{ds^2}+\tensor{\Gamma}{^{\rho}_{\mu\lambda}}\frac{\tensor{dx}{^\mu}}{ds}\frac{\tensor{dx}{^\lambda}}{ds} =\frac{\tensor{h}{^{\nu\rho}}}{p+\rho}\left(\tensor{J}{_\nu}-\tensor{\nabla}{_\nu} p \right).
\end{equation}
From Eq.~(\ref{eqmot}) it follows that the quantity
\begin{equation}
\frac{\tensor{h}{^{\nu\rho}}}{p+\rho}\left(\tensor{J}{_\nu}-\tensor{\nabla}{_\nu} p \right) ,
 \end{equation}
determines the deviation of a particle's worldline from the geodesic motion, and thus represents a generalized force  $\tensor{\mathcal{F}}{^\rho}$ that makes the motion nongeodesic
\begin{equation}
    \tensor{\mathcal{F}}{^\rho}=\frac{\tensor{h}{^{\nu\rho}}}{p+\rho}\left(\tensor{J}{_\nu}-\tensor{\nabla}{_\nu} p \right).
\end{equation}
Evaluating Eq.~(\ref{EMdiv}) for a perfect fluid with $\mathcal{L}_{m}=p$, the generalized force becomes
\begin{eqnarray}\label{force}
    \tensor{\mathcal{F}}{^\rho} &=& \frac{2\tensor{h}{^{\nu\rho}}\tensor{\nabla}{^\mu} \tensor{H}{_{\mu\nu}} +\tensor{f}{_T} \tensor{D}{^\rho} (\rho-3p) - \tensor{D}{^\rho}( pF)}{(p+\rho)(F+2\tensor{f}{_T})},
\end{eqnarray}
where we have defined the total derivative along the flow
\begin{equation} 
\tensor{D}{^\mu} \equiv \tensor{\nabla}{^\mu} +\tensor{u}{^\mu}\frac{d}{ds} .
\end{equation}

\subsection{Newtonian Limit}

The Newtonian limit is obtained in the weak field and small velocities approximation, $\vec{v}^2 \ll c^2$, where $\vec{v}$ is the three-dimensional velocity of a massive test particle. This corresponds to approximating the metric as \footnote{We use Latin indices to represent the values 1,2,3.}  
\begin{equation}
ds^2 = -(1+2\Phi) dt^2 + \tensor{\delta}{_{ij}} \tensor{dx}{^i} \tensor{dx}{^j} ,
\end{equation}
with $\Phi$ being the Newtonian potential and  $\delta_{ij}$ being the Kronecker delta.

Moreover, we neglect the pressure term in the expression of the energy-momentum tensor, by assuming $\rho \gg p$. Then the only non-zero component of the matter energy-momentum tensor is given by $\tensor{T}{_{00}}=\rho$.

For the function $f(R,T)$ we assume that it can be approximated by a first order Taylor expansion around a fixed point $\left(\tensor{R}{_0},\tensor{T}{_0}\right)$,
\begin{eqnarray}
f(R,T)&=&f\left (\tensor{R}{_0},\tensor{T}{_0}\right)+\left(R-\tensor{R}{_0}\right) \tensor{f}{_R}\left(\tensor{R}{_0},\tensor{T}{_0}\right)\nonumber\\
&&+\left(T-\tensor{T}{_0}\right) \tensor{f}{_T}\left( \tensor{R}{_0},\tensor{T}{_0}\right)+ \cdots ,
\end{eqnarray}
which implies that $\tensor{f}{_R}\approx \tensor{f}{_R}\left(\tensor{R}{_0},\tensor{T}{_0}\right)={\rm constant}$, and $\tensor{f}{_T}\approx \tensor{f}{_T}\left(\tensor{R}{_0},\tensor{T}{_0}\right)={\rm constant}$, respectively. 

\subsubsection{The generalized Poisson equation}

To obtain the generalized Poisson equation in the Newtonian limit of  Herglotz $f(R,T)$ gravity we consider the $00$  component of Eq.~(\ref{tracelessform}). Noting that $\tensor{\Theta}{_{00}} =-\Theta= -2\rho$ and working to first order in the potential $\Phi$, we find 
\begin{eqnarray}
       \Delta\Phi = \frac{1}{2f_R}\left(f_T+\frac{F}{2}\right)\rho -\frac{2}{3f_R}\left(H_{00}-\frac{1}{4}g_{00}H\right) .
\end{eqnarray}

In the limiting case of general relativity, where $F=16\pi$, $\tensor{f}{_R}=1$, $\tensor{f}{_T}=0$, $\tensor{H}{_{\mu \nu}}=0$, the standard Poisson equation of Newtonian gravity,  $\Delta \Phi =4\pi \rho$, is recovered. In the Newtonian limit with constant $f_{R}$, the Herglotz tensor takes the simple form $\tensor{H}{_\mu _\nu}=\tensor{f}{_R} \tensor{K}{_\mu _\nu}$. Defining an effective coupling constant $\tensor{k}{^2_{eff}}=\frac{\tensor{f}{_T}+8 \pi}{2 \tensor{f}{_R}}$  , we obtain, to first order in the Herglotz field
\begin{eqnarray} 
    \Delta \Phi = k_{eff}^2\rho -\frac{1}{6}\left(1+2\Phi\right)\partial_i \lambda^i+ \frac{2}{3}\lambda ^i \partial_i\Phi . 
\end{eqnarray}
    Considering spherical symmetry of the Herglotz vector $\tensor{\lambda}{^\mu} = \tensor{\lambda}{_\mu} = \left(0,\vec{\psi} \right) = (0,\psi(r) ,0,0)$, the generalized Poisson equation reads
    \begin{eqnarray}
        \Delta \Phi &=&   \tensor{k}{^2_{eff}}\rho-\frac{1}{6}\left(1+2\Phi\right)\nabla\cdot \vec\psi+ \frac{2}{3}\vec\psi \cdot\nabla\Phi \nonumber\\
        &=& \tensor{k}{^2_{eff}}\rho -(1+2\Phi)\frac{1}{6r^2}\frac{d}{dr}(r^2\psi)+ \frac{2}{3}\psi \frac{d\Phi}{dr},
    \end{eqnarray} 
    or in the vacuum with $\rho=0$,
    \begin{equation} \label{PoissonVac}
\frac{1}{r}\frac{d^2}{dr^2}(r\Phi) = -(1+2\Phi)\frac{1}{6r^2}\frac{d}{dr}(r^2\psi)+ \frac{2}{3}\psi \frac{d\Phi}{dr} .
    \end{equation}

By introducing a new variable $u=r\Phi$, and taking into account that 
\begin{equation}
r\frac{d\Phi}{dr}=\frac{du}{dr}-\frac{u}{r},
\end{equation}
Eq.~(\ref{PoissonVac}) takes the form
\begin{equation}\label{u}
\frac{d^2u}{dr^2}-\frac{2}{3}\psi \frac{du}{dr}+\frac{2}{3}\psi \frac{u}{r}=-\frac{1}{6}\left(r+2u\right)\frac{1}{r^2}\frac{d}{dr}\left(r^2\psi\right).
\end{equation}
To explore possible modifications, we consider specific functional forms of $\psi(r)$. \\
(a) In the case $\psi=-\frac{3\psi_0}{2}={\rm constant}$, Eq.~(\ref{u}) becomes
\begin{equation}\label{u1}
    \frac{d^2u}{dr^2}+\psi_0 \frac{du}{dr}-2 \psi_0 \frac{u}{r}=\frac{1}{2} \psi_0.
\end{equation}
For large values of $r$ we can neglect the third term in Eq.~(\ref{u1}) proportional to $1/r$. Hence, in the constant $\psi$ case we obtain for $u$ the equation
\begin{equation}
\frac{d^2u}{dr^2}+ \tensor{\psi}{ _0}\frac{du}{dr}=\frac{1}{2} \tensor{\psi}{_0},
\end{equation}
giving for the modified Newtonian potential the expression
\begin{equation}
\Phi (r)=\frac{\tensor{C}{_2}}{r}+\frac{\tensor{C}{_1}}{\tensor{\psi}{_0}}\frac{e^{-\tensor{\psi}{_0}r}}{r}+\frac{1}{2}.
\end{equation}
The constant term in the potential can be neglected since it does not give any contribution to the gravitational force and by analogy with Newtonian gravity we can choose the constants $\tensor{C}{_2}=-GM, \tensor{C}{_1} = 0$, or $\tensor{C}{_2}=0, \tensor{C}{_1} = -GM \tensor{\psi}{_0}$ where $M$ is the mass of the central object. The first choice of constants gives exactly the Newtonian potential with no modifications and the second choice gives the series expansion 
\begin{equation} \label{ModPhi1}
\Phi(r) \approx -\frac{GM}{r}+GM \tensor{\psi}{_0}- \frac{GM\psi_0^2}{2}r + \frac{GM\psi_0^3}{6}r^2+\mathcal{O}\left(r\right)^3 .
\end{equation}
Since the correction terms to the Newtonian potential grow with $r$, we discard this case.

(b) The case $\psi=\frac{3 \tensor{\psi}{_0}}{2r}$. For this choice of $\psi$, Eq.~(\ref{u}) takes the form
\begin{equation}
    \frac{d^2 u}{dr^2} - \frac{\psi_0}{r} \frac{du}{dr} + \frac{3}{2} \frac{\psi_0}{r^2} u= -\frac{\psi_0}{4r}.
\end{equation}
The general solution of the above equation can be given as
\begin{eqnarray}
   u(r)&=&  C_1 r^{\frac{1}{2} \
\left(\psi_0-\sqrt{\psi_0+\frac{1}{\psi_0}-4} \sqrt{\psi_0}+1\right)} \nonumber\\
&&+C_2 r^{\frac{1}{2} \
\left(\psi_0+\sqrt{\psi_0+\frac{1}{\psi_0}-4} \sqrt{\psi_0}+1\right)}-\frac{r}{2}.
\end{eqnarray}
Since the Newtonian term $\Phi \approx GM/r$ cannot be reproduced from the above solution for any real value of $\tensor{\psi}{_0}$ we will discard this case.  

(c) For $\psi= \frac{3\tensor{\psi}{_0}}{2r^2}$, Eq.~(\ref{u}) becomes
\begin{equation}
\frac{d^2u}{dr^2}-\frac{\tensor{\psi}{_0}}{r^2}\frac{du}{dr}+\frac{\tensor{\psi}{_0}}{r^3}u=0,
\end{equation}  
leading to
\begin{equation}
\Phi(r)=\tensor{C}{_1}+\frac{\tensor{C}{_2}}{\tensor{\psi}{_0}}e^{-\tensor{\psi}{_0}/r}.
\end{equation}
In principle, it is possible to neglect the $C_1$ integration constant, as it is not physically measurable. However, for mathematical consistency, we keep  the constant and take the series expansion of the potential for large values of $r$, obtaining
\begin{equation}\label{pot}
\Phi(r)= C_1 + \frac{C_2}{\psi _0}+  C_2\left[-\frac{1}{r}+\frac{\psi _0}{2r^2}-\frac{\psi_0^2}{6r^3}+\mathcal{O}\left(\frac{1}{r}\right)^4\right].
\end{equation}
From Eq.~(\ref{pot}) it follows immediately that $\tensor{C}{_2}=GM$ and $C_1 = -\frac{GM}{\psi_0}$ leads to the correct first order Newtonian approximation, and hence we get the corrected Newtonian potential up to fourth order  as
\begin{equation} \label{ModPhi2}
\Phi(r) = -\frac{GM}{r} +\frac{GM\psi_0}{2}\frac{1}{r^2}-\frac{GM\psi_0^2}{6}\frac{1}{r^3} +\mathcal{O}\left(\frac{1}{r}\right)^4.
\end{equation}

\subsubsection{The equation of motion of massive particles}

To find the equation of motion for a massive test particle we can evaluate Eq. (\ref{eqmot}) in the Newtonian limit of the gravitating system. In this case $\tensor{D}{^i} \rightarrow \tensor{\nabla}{^i}$ and $\tensor{h}{^{\mu i}} \rightarrow \tensor{g}{^{\mu i}}$ so that Eq. (\ref{force}) reads
\begin{eqnarray} 
\tensor{\mathcal{F}}{^i} &=&\frac{\tensor{f}{_T}}{2\left(8\pi+\tensor{f}{_T}\right)\rho}\tensor{\delta}{^{ij}} \tensor{\partial}{_j}\rho +\frac{2 \tensor{f}{_R}\tensor{\lambda}{^k}}{\left(8\pi+\tensor{f}{_T} \right)\rho}\tensor{\delta}{^{ij}} \tensor{\partial}{_k} \tensor{\partial}{_j}\Phi \nonumber\\
&& + \frac{\tensor{f}{_R} \tensor{\partial}{_k} \tensor{\partial}{^k} \Phi }{\left(8\pi+\tensor{f}{_T} \right)\rho} \tensor{\delta}{^{ij}} \tensor{\lambda}{_j} + \frac{2 \tensor{f}{_R} \tensor{\partial}{^k}\Phi}{\left(8\pi+\tensor{f}{_T} \right)\rho}\tensor{\delta}{^{ij}} \tensor{\partial}{_j} \tensor{\lambda}{_k},
\end{eqnarray}
where we used the fact that $f_R$ is constant and the fluid is pressureless.

Expressing the extra acceleration in terms of the familiar Newtonian equation of motion we have 
\begin{equation}
\vec{a} = \tensor{\vec{a}}{_N} + \tensor{\vec{a}}{_E} \; ,
\end{equation}
where $\tensor{\vec{a}}{_N} $ is the usual Newtonian acceleration and
\begin{eqnarray} 
    \tensor{\vec{a}}{_E} &=& \frac{\tensor{f}{_T}}{2\left(8\pi+\tensor{f}{_T} \right)\rho}\nabla\rho + \frac{2 \tensor{f}{_R}}{\left(8\pi+ \tensor{f}{_T} \right)\rho}\left(\vec\psi \cdot \vec{\nabla} \right)\nabla\Phi \nonumber\\
    &&+ \frac{\tensor{f}{_R} \Delta\Phi}{\left(8\pi+\tensor{f}{_T} \right)\rho}\vec{\psi} + \frac{2 \tensor{f}{_R}}{\left(8\pi +\tensor{f}{_T} \right)\rho}\nabla\Phi\cdot \vec{\nabla} \vec\psi  .
\end{eqnarray}
Equivalently, in spherical symmetry this becomes
\begin{eqnarray}
 \tensor{a}{_E} \hat{r} &=& \frac{\tensor{f}{_T}}{2\left(8\pi +\tensor{f}{_T} \right)\rho} \frac{d\rho}{dr} \hat{r} \nonumber\\
&&+\frac{\tensor{f}{_R}}{(8\pi +\tensor{f}{_T}) \rho}\left(3\psi\frac{d^2\Phi}{dr^2}+\frac{2}{r}\frac{d\Phi}{dr}\frac{d}{dr}(r\psi)\right)\hat{r} .
\end{eqnarray}
Alternatively, once the modified potential $\Phi$ is known, the additional acceleration can be obtained directly from the deviation of $\Phi$ relative to the Newtonian form. For the potential given in Eq.~(\ref{ModPhi2}), we get
\begin{eqnarray}
\nabla\Phi = \frac{GM}{r^2} -\frac{GM \psi_0}{r^3}+ \frac{GM\psi_0^2}{2r^4} +\mathcal{O}\left(\frac{1}{r}\right)^5,
\end{eqnarray}
from which we approximate the extra acceleration as
\begin{equation}\label{a_E2}
\tensor{a}{_E} = \frac{GM \tensor{\psi}{_0}}{r^3} \left( \frac{\tensor{\psi}{_0}}{2r}-1\right) .
\end{equation}

\subsubsection{The perihelion precession}

Using the expression for the additional acceleration, we can estimate its contributions to the perihelion precession of planets as well as to the deflection of light. For a system with central mass $M$ the Laplace-Runge-Lenz vector (LRL) is given by \cite{Brill1999} 
\begin{equation}
\vec{A} = \vec v \times \vec L- \epsilon GM\hat{r} ,
\end{equation}
where $\vec{v} $ is the velocity and $\vec L$ is the specific angular momentum. Also, $\epsilon$ parametrizes time-like and null geodesics, hence $\epsilon=1$ for massive particles and $\epsilon=0$ for light. Classically, the LRL vector is conserved in a central potential $V(r)$, thus the change in perihelion precession $\Delta\phi$ over one orbit can be obtained by the derivative of $\vec A$ with respect to the polar angle $\theta$,
\begin{equation}
\frac{d \vec A}{d\theta} = \frac{d\vec A}{dt}\left(\frac{d\theta}{dt}\right)^{-1}  = r^2 \left(\frac{\partial V(r)}{\partial r}+ \tensor{a}{_E}-\frac{  \epsilon GM}{r^2} \right)\hat{\theta} ,
\end{equation}
where $\tensor{a}{_E}$ is the magnitude of the extra acceleration. If we consider a general LRL vector $\vec A = A\hat{x}$, we arrive at the relation
\begin{equation} \label{orbit}
\frac{1}{r} = \frac{1}{L^2} \left(  \epsilon GM + A\cos\theta\right) .
\end{equation}
Then by using the rate of change of direction for any vector 
\begin{equation}
\vec\omega = \frac{\vec A \times \dot{\vec A}}{A^2} ,
\end{equation}
and the post-Newtonian (Schwarz{s}child) potential \cite{Carroll_2019}
\begin{equation}\label{Potential}
\begin{aligned}
V(r) &= -\frac{ \epsilon GM }{r} -\frac{GM L^2}{r^3},
\end{aligned}
\end{equation}
we obtain for the shift in the massive case ($\epsilon =1$)
\begin{eqnarray}
\Delta \phi &=& \int_0^{2\pi} \omega\,dt \nonumber\\
&=& \int_0^{2\pi}\frac{1}{A^2} \left|\vec A\times \frac{d\vec A}{d\theta}\right| d\theta \nonumber \\
&=& \frac{1}{A}\int_0^{2\pi}   \left(\frac{3GML^2}{r^2} +  r^2 \tensor{a}{_E}\right) \left|\hat{x} \times\hat{\theta}\right| \, d\theta \nonumber \\
&=& \frac{6\pi GM}{a(1-e^2)} + \frac{ L^4}{A}\int_0^{2\pi} \frac{\tensor{a}{_E}\cos\theta\,d\theta}{( GM+A\cos\theta)^2} \nonumber \\
&\equiv& \tensor{\Delta \phi}{_{GR}}+\tensor{\Delta\phi}{_E} ,
\end{eqnarray}
since $GM/L^2 = 1/a(1-e^2)$ for eccentricity $e$ and semi-major axis $a$.

We can put bounds on the Herglotz field by considering Eq. (\ref{a_E2}) so that 
\begin{eqnarray}
\tensor{\Delta \phi}{_E} &=& \frac{ L^4}{A}\int_0^{2\pi} \frac{\tensor{a}{_E}\cos\theta\,d\theta}{( GM+A\cos\theta)^2}\nonumber \\
&=&\frac{GM \psi_0 }{A}\int_0^{2\pi}\bigg(\frac{\psi_0}{2L^4}( GM +A\cos\theta)^2 \nonumber \\
&&\hspace{20mm}-\frac{1}{L^2}( GM+A\cos\theta)\bigg) \cos \theta d\theta \nonumber \\
&=&\frac{\pi \psi_0^2}{a^2(1-e^2)^2}  -\frac{\pi \psi_0}{a(1-e^2)} .
\end{eqnarray}
Then, for the planet Mercury we have $a = 57. 91\times 10^{11}$ cm and $e=0.205615$, which means that given the observed value of the precession $\tensor{\Delta \phi} {_{obs}}= 43.11 \pm 0.21$ arcsec per century \cite{French1968SpecialRelativity}, any additional precession must obey $\tensor{\Delta \phi}{_E} \leq \tensor{\Delta\phi}{_{obs}}-\tensor{\Delta\phi}{_{GR}} = 0.1^{+0.2}_{-0.1} $ arcsec per century. If we take the the extra precession to contain the entire deviation from observation then this gives an upper bound of $\tensor{\psi}{_0} \approx 5.5\times 10^{12} $ cm for the case of an inverse square Herglotz function. 

\subsubsection{The deflection of light}

To calculate the deflection of light we take $\epsilon=0$ and consider the deflection bounds $-\pi/2 $ to $\pi/2$. Doing this gives 
\begin{eqnarray}
\Delta \phi &=& \int_{-\pi/2}^{\pi/2} \omega\,dt \nonumber \\
&=& \int_{-\pi/2}^{\pi/2} \left[\frac{3GM A  \cos^2\theta}{L^2}+ \frac{ L^4 \tensor{a}{_E}} { A^3\cos^2\theta}\right]\cos\theta\,d\theta \nonumber \\
&=& \frac{4 GM}{b} + \frac{b^2}{A}\int_{-\pi/2}^{\pi/2} \frac{\tensor{a}{_E}}{\cos\theta} d\theta \label{philight},
\end{eqnarray}
by defining the impact parameter $b= L^2/A$. Looking at Eq. (\ref{philight}) we see that the extra deflection of photons depends on the magnitude of the LRL vector. In the case of photons we will assume that they have an effective mass given by $m_{eff}=h\nu /c^2$, where $h$ is Planck's constant, and $\nu$ is the frequency, a relation which is obtained by assuming that the photon energy $E=h\nu$ is related to an effective mass via Einstein's relation $E=m_{eff}c^2$.  Thus for the  magnitude of the Laplace-Runge-Lenz vector we obtain $A = h_p^2 \nu^2 b$ where $h_p=h/m_p$ is the Planck constant divided by the Planck mass $m_p$. Then using Eq. (\ref{a_E2}) we find
\begin{eqnarray}
\tensor{\Delta \phi}{_E} &=& \frac{b^2GM\psi_0}{A}\int_{-\pi/2}^{\pi/2}\left(\frac{\psi_0 }{2b^4}\cos^3\theta -\frac{1}{b^3}\cos^2\theta\right) d\theta   \nonumber \\
&=& \frac{2GM\psi_0^2}{3Ab^{2}} -\frac{\pi GM\psi_0}{2Ab} \nonumber\\
&=& \frac{GM\psi_0 }{h_p^2\nu^2b^2}\left(\frac{2\psi_0}{3b}-\frac{\pi}{2}\right) \label{light1}.
\end{eqnarray}

Thus, we find that in the presence of the Herglotz field, the bending of light depends on the frequency. In particular, the magnitude of the bending is proportional to the square of the wavelength. Interestingly enough, this result matches the scaling proposed by light traveling through a plasma medium \cite{Slava2019,Clegg_1998,perlick2000}. 

For the specific case of impact parameter being equivalent to the solar radius, $b = R_\odot$, we find
\begin{equation} \label{plasangle}
    \frac{\tensor{\Delta\phi}{_p}}{\tensor{\Delta \phi}{_{GR}}} = 1.02\times 10^{-7}\left(\frac{\lambda}{\mu\text{m}}\right)^2 ,
\end{equation}
where $\lambda$ is the wavelength of the light and $\tensor{\Delta\phi}{_p}$ is the so called plasma deflection angle predicted in \cite{Slava2019} which agrees with recent observations using the Cassini spacecraft \cite{Bertotti2003}. Using $\tensor{\Delta\phi}{_p} =\tensor{\Delta\phi}{_E}$ and the above relation we find 
\begin{equation}
    \frac{\psi_0 }{2h_p^2b}\left(\frac{\psi_0}{3b}-\frac{\pi}{4}\right) = 1.02\times 10^{-7}\left(\frac{1}{\mu\text{m}}\right)^2 ,
\end{equation}
when considering Eq. (\ref{light1}). Taking the positive solution we find $\psi_0 \approx 1.6\times 10 ^{11}$ cm. This value falls within the upper bound of that calculated via the perihelion shift of Mercury, and is only one order of magnitude short of the assumption that the Herglotz field accounts for the entire precession difference not described by GR. 

\section{Cosmological applications}\label{section4}

To explore the cosmological implications of Herglotz-type $f(R,T)$ gravity, we assume an isotropic and  homogeneous flat Friedmann-Lemaitre-Robertson-Walker (FLRW) metric of the form
\begin{align}
    ds^2 = -dt^2 +a^2(t)\delta_{ij}dx^idx^j ,
\end{align}
where $a(t)$ is the scale factor and $t$ denotes the comoving time. By spatial homogeneity, the Herglotz field is constrained to the form
\begin{align}
    \lambda_\mu = (\phi(t),0,0,0),
\end{align}
with $\phi(t)$ a smooth function of the cosmic time. Similarly, we must have that $f(R,T)$ and its subsequent derivatives only depend on $t$. Finally, we model the matter content as a perfect fluid with pressure $p$ and energy density $\rho$,
\begin{align}\label{T}
    T_{\mu\nu} = (p+\rho)u_\mu u_\nu+pg_{\mu\nu} ,
\end{align}
where the fluid four-velocity obeys $u_\mu u^\mu = -1$. For the matter Lagrangian, we adopt the common choice $\mathcal{L}_{m}=p$. 

\subsection{The generalized Friedmann equations}

By taking the typical value $F=16\pi$, the FLRW equations for Herglotz-type $f(R,T)$ gravity read
\begin{align}
\begin{split}
        &3f_R\dot{H} + 3\tensor{f}{_R} H^2 + 3\left(\phi \tensor{f}{_R} - \tensor{\dot{f}}{_R} \right)H -\frac{1}{2}f\\
    &= -\tensor{f}{_T}(\rho+p)-8\pi\rho ,\\
\end{split}\\\begin{split}
     &2\tensor{f}{_R}\dot{H} +\left(\phi \tensor{f}{_R}-\tensor{\dot{f}}{_R} \right)H +\tensor{\ddot{f}}{_R}-\dot{\phi}\tensor{f}{_R} -2\phi \tensor{\dot{f}}{_R}+\phi^2 \tensor{f}{_R}\\
     &= -\left(8\pi+\tensor{f}{_T} \right)(\rho+p) ,
\end{split}
\end{align}
where $H\equiv \dot{a}/a$ and an overhead dot denotes differentiation with respect to $t$. These equations reduce to the standard $f(R,T)$ results \cite{farias2021}  in the absence of the Herglotz field $(\phi=0)$
\begin{align}
    3\left(\tensor{f}{_R}\dot{H} + \tensor{f}{_R} H^2 -\tensor{\dot{f}}{_R}H \right) -\frac{1}{2}f&= -\tensor{f}{_T}(\rho+p)-8\pi\rho ,\\
    2\tensor{f}{_R}\dot{H} \tensor{-\dot{f}}{_R}H +\tensor{\ddot{f}}{_R} &=-(8\pi+f_T)(\rho+p) ,
\end{align}
and to the Friedmann equations of nonconservative gravity \cite{Paiva_2022}, when $f(R,T)=R$ 
\begin{align}\label{Frn}
     3H^2 - 3\phi H &= 8\pi\rho ,\\
    2\dot{H} +\phi H -\dot{\phi} +\phi^2&=-8\pi(\rho+p) ,
\end{align}
 It will also prove useful to write the FLRW equations in an equivalent form
\begin{align}
   3 H^2 &= \frac{8\pi}{\tensor{f}{_R}}\left(\rho+\tensor{\rho}{_{eff}} \right) ,\label{Fr1}\\
    2\dot{H} +3H^2 &= -\frac{8\pi}{\tensor{f}{_R}}\left(p+\tensor{p}{_{eff}} \right) , \label{Fr2}
\end{align}
using an effective density and pressure,
\begin{align}
    \tensor{\rho}{_{eff}} &= \frac{1}{8\pi}\left(3\left(\phi \tensor{f}{_R}-\tensor{\dot{f}}{_R} \right)H +\frac{1}{2}\left(\tensor{f}{_R} R-f \right)+\tensor{f}{_T}(\rho+p)\right)\label{rhoeff} , \\
    \tensor{p}{_{eff}} &= \frac{1}{8\pi}\bigg( \tensor{\ddot{f}}{_R}-\frac{1}{2}\left(\tensor{f}{_R} R-f \right)-2\left(\phi \tensor{f}{_R}-\tensor{\dot{f}}{_R} \right)H -\dot{\phi} \tensor{f}{_R} \nonumber\\
    &-2\phi \tensor{\dot{f}}{_R} +\phi^2 \tensor{f}{_R}\bigg) \label{peff} .
\end{align}

By eliminating the $3H^2$ term between Eqs.~(\ref{Fr1}) and (\ref{Fr2}) we obtain the equation describing the dynamical evolution of $H$ as
\begin{equation}\label{62}
\dot{H}=-\frac{4\pi}{\tensor{f}{_R}}\left(\rho+p+\tensor{\rho}{_{eff}}+\tensor{p}{_{eff}}\right).
\end{equation}

By taking the time derivative of Eq.~(\ref{Fr1}), and with the use of Eq.~(\ref{62}), we obtain the generalized matter conservation equation in Herglotz type $f(R,T)$ gravity as
\begin{eqnarray}
\dot{\rho}+3H(\rho+p)&=&-\left[\tensor{\dot{\rho}}{_{eff}}+3H\left(\tensor{\rho}{_{eff}}+\tensor{p}{_{eff}}\right)\right]\nonumber\\
&&-\left(\rho+\tensor{\rho}{_{eff}}\right)f_R\frac{d}{dt}\frac{1}{\tensor{f}{_R}}.
\end{eqnarray}
In the cosmological framework this is explicitly evaluated as the time component of $\tensor{J}{_\nu}$ in Eq. (\ref{EMdiv})
\begin{eqnarray}
 \tensor{J}{_0} =(\rho+p) \Gamma &=&-\frac{1}{8\pi+f_T}\Bigg[3\phi\left(\dot{H}+H^2+\phi H\right) \tensor{f}{_R}  \nonumber \\
  &&-3\phi H \tensor{\dot{f}}{_R}+ (\rho+p)\tensor{\dot{f}}{_T} + \frac{1}{2}\left(\dot{\rho} -\dot{p}\right)\tensor{f}{_T}\Bigg] . \nonumber \\
 \end{eqnarray}

Physically, $\Gamma$ can be interpreted as the rate of energy creation or annihilation in the cosmic environment, arising from the dissipative effects associated with the geometry-matter coupling and the Herglotz field.

As an indicator of the dynamical evolution of the Universe we introduce the deceleration parameter $q$, defined as
\begin{equation}
q\equiv\frac{d}{dt}\frac{1}{H}-1.
\end{equation} 

With the use of Eqs.~(\ref{62}) and (\ref{Fr1}) the deceleration parameter can be expressed as
\begin{equation}
q=\frac{3}{2}\frac{\rho+p+\tensor{\rho}{_{eff}}+\tensor{p}{_{eff}}}{\rho+\tensor{\rho}{_{eff}}}-1.
\end{equation}

The accelerated expansion corresponds to negative values of $q$, $q<0$. Hence the condition for the {accelerating} expansion in Herglotz-type $f(R,T)$ gravity reads
\begin{equation}
\tensor{\rho}{_{eff}}+3\tensor{p}{_{eff}}+\rho+3p<0.
\end{equation}  

In the particular case of a pressureless fluid (dust) we have $p=0$, and the deceleration parameter becomes
\begin{eqnarray}
q &=& \frac{3}{2}\frac{\left(8\pi+\tensor{f}{_T} \right)\rho +\tensor{\ddot{f}}{_R}-(2\phi+H)\tensor{\dot{f}}{_R}-(\dot{\phi} +\phi^2-\phi H)\tensor{f}{_R}}{\left(8\pi+\tensor{f}{_T} \right)\rho +3\phi H \tensor{f}{_R}-3H \tensor{\dot{f}}{_R} +(\tensor{f}{_R}R-f)/2}\nonumber
\\&&-1.
\end{eqnarray}
The acceleration condition is then given as
\begin{eqnarray}
 \hspace{-0.5cm}&&\tensor{f}{_R}R -f -\left(8\pi +\tensor{f}{_T} \right)\rho \nonumber>\\
\hspace{-0.5cm}&&3\left(\tensor{\ddot{f}}{_R}-2\phi \tensor{\dot{f}}{_R}+H \tensor{\dot{f}}{_R}-\dot{\phi}\tensor{f}{_R}+\phi^2\tensor{f}{_R}-\phi H \tensor{f}{_R} \right) .
\end{eqnarray}
 
\subsection{Specific cosmological models} 

In the models considered below, we assume the Universe is filled with a pressureless fluid (dust), setting $p=0$. Under this assumption, the trace of the energy-momentum tensor reduces to $T=-\rho$.
 
\subsubsection{Model I: $f(R,T)=R+\alpha T$}

We begin by analyzing one of the most simple extensions to general relativity, choosing $f(R,T)=R+\alpha T$ where $\alpha\in \mathbb{R}$. In this case the FLRW equations read
\begin{align}\label{Model1constraint}
    3H^2 &= 3\phi H + \frac{1}{2}(16\pi +3\alpha)\rho , \\
    2\dot{H}+3H^2  &=\dot{\phi}+2\phi H -\phi^2+\frac{\alpha}{2} \rho  .
\end{align}
Eq. (\ref{Model1constraint}) is an algebraic constraint, which can be used to isolate the density function. Replacing this in the dynamical equation gives 
\begin{equation}\label{Model1dynamic}
    2\dot{H}-\dot{\phi} = 2\phi H-3H^2 -\phi^2+\frac{3\alpha}{16\pi +3\alpha}(H^2-\phi H) ,
\end{equation}
or, equivalently,
\begin{equation}
\dot{H}=\frac{\dot{\phi}}{2}-\frac{1}{2}\phi H-\frac{\phi^2}{2}-\frac{1}{2}\left(\alpha +8\pi\right)\rho.
\end{equation}

The deceleration parameter is obtained as
\begin{equation}
q=-\frac{3}{2}\left[\frac{\dot{\phi}-\phi H-\phi^2-(\alpha +8\pi)\rho}{3\phi H+\left(16\pi+3\alpha\right)\rho /2}\right]-1.
\end{equation}
The condition for accelerated expansion gives for the matter density the constraint
\begin{equation}
\frac{8\pi \rho}{3}<\dot{\phi}-\phi^2+\phi H. 
\end{equation}

The dynamical equations involve two unknown functions, so an additional constraint is required to close the system. It is also convenient to introduce a set of dimensionless variables,
\begin{equation}\label{dimlessparams}
     \tau=H_0 t,~ H=H_0 h,~ \rho=\frac{3H_0^2}{8 \pi} r, ~\phi=H_0 \Phi, ~ \alpha=\frac{16\pi}{3}  A ,
 \end{equation}
 where $H_0$ represents the present-day value of the Hubble constant. Thus, we have the dynamical equation  
\begin{eqnarray}\label{Model1dim}
    2 \frac{dh(\tau)}{d\tau} - \frac{d\Phi(\tau)}{d\tau} &=&\frac{2+A}{1+A}\Phi(\tau)h(\tau)- \Phi^2(\tau)\nonumber\\
    &&-\frac{3+2A}{1+A}h^2(\tau),
\end{eqnarray}
supplemented by the algebraic constraint
\begin{equation}\label{Constrainmodel1}
    h^2(t)=\Phi(\tau) h(\tau) + (1+A) r(\tau).
\end{equation}

\paragraph{Analytical solutions.} When considering a simple linear model, there are two ways to obtain analytical solutions. These occur when either the Hubble or the Herglotz functions are a constant, that is $h(\tau)=h_0$ or $\Phi(\tau)=\Phi_0$ with $h_0,\Phi_0 \in \mathbb{R}$. The former case corresponds to a de Sitter type solution, whereas the latter case can be interpreted as a situation in which a constant is a source of the non-conservation of energy and momentum.

Aiming to obtain a de Sitter solution, Eq. (\ref{Model1dim}) reads 
\begin{equation}\label{Analyticdifphi}
  \frac{d\Phi(\tau)}{d\tau} = \Phi^2(\tau) - \frac{(2+A)h_0}{1+A}\Phi(\tau) + \frac{(3+2A)h_0^2}{1+A} .
\end{equation}
This is an integrable differential equation, and it is a particular form of the Riccati equation.  In order to simplify the equation we introduce the notations $\sigma = (2+A)h_0/(2+2A)$, and $\psi=(3+2A)h_0^2/(1+A)$, respectively, which allow us to write it as
\begin{equation}
 \frac{d\Phi(\tau)}{d\tau} = \Phi^2(\tau) - 2\sigma\Phi(\tau) + \psi ,
 \end{equation}
which, in turn, yields
\begin{equation}\label{LinearODE}
\frac{d^2u}{d\tau^2}+2\sigma\frac{du}{d\tau}+\psi u = 0 ,
\end{equation}
by using the transformation $\Phi = -d(\ln u)/d\tau$. This is a simple second order ODE, where the solutions depend on the sign of the discriminant $\Delta =4\left(\sigma ^2-\psi\right)=-\left(8+16A+7A^2\right)h_0^2/(1+A)^2, ~A\neq -1$. Taking cases on $\Delta$ the solutions for Eq. (\ref{LinearODE}) can be combined with the transformation to obtain $\Phi$. If $\Delta >0$,
then 
\begin{eqnarray}
\Phi(\tau)&=&\sigma -\frac{\sqrt{\Delta}}{2}\tanh \left[\frac{\sqrt{\Delta}}{2}\left(\tau-\tau_0\right)\right], \nonumber\\
A&\in & \left(-\frac{\left(4+\sqrt{2}\right)}{7}, \frac{\left(-4+\sqrt{2}\right)}{7}\right).
\end{eqnarray}  

For $\Delta <0$, the solution of Eq.~(\ref{Analyticdifphi}) is given by
\begin{eqnarray}
\Phi (\tau)&=&\sigma +\frac{\sqrt{-\Delta}}{2}\tan \left[\frac{\sqrt{-\Delta}}{2}\left(\tau -\tau_0\right)\right], \nonumber\\
A&\in & \left(-\infty, -\frac{\left(4+\sqrt{2}\right)}{7}\right)\bigcup \left(\frac{\left(-4+\sqrt{2}\right)}{7},+\infty\right).\nonumber\\
\end{eqnarray}
Finally, for the case $\Delta=0$, we obtain
\begin{equation}
\Phi (\tau)=\left(1\mp \sqrt{2}\right)h_0-\frac{1}{\tau-\tau_0}, ~A=\frac{2}{7}\left(-4\mp \sqrt{2}\right).
\end{equation} 

In contrast, if we consider when $\Phi(\tau)=\Phi_0$, then (\ref{Model1dim}) reads 
\begin{equation}
    \frac{d h(\tau)}{d\tau} =-\frac{3+2A}{2+2A}h^2(\tau)+\frac{(2+A)\Phi_0 }{2+2A}h(\tau) -\frac{\Phi_0^2}{2}  ,
\end{equation}
which reduces to exactly Eq. (\ref{LinearODE}) with the transformation $ (3+2A)h = (2+2A) d(\ln u)/d\tau$, when $\sigma = -(2+A)\Phi_0/(4+4A)$, and $\psi = (3+2A)\Phi_0^2/(4+4A)$. In this case the discriminant reads $\Delta = \left(4+3A\right)^2\Phi_0^2/(2 + 2A)^2, ~A\neq-1$. Under the new transformation if $\Delta >0$ we get 
\begin{eqnarray} 
h(\tau)&=& \frac{2+2A}{3+2A}\left(\frac{\sqrt{\Delta}}{2}\tanh \left[\frac{\sqrt{\Delta}}{2}\left(\tau-\tau_0\right)\right]-\sigma \right), \nonumber\\
A&\in & \left(-\infty, -\frac{4}{3}\right)\bigcup \left(-\frac{4}{3}, -1\right) \bigcup \left(-1,+\infty\right).
\end{eqnarray}  
Now, since $\Delta$ is positive definite there are no solutions for $\Delta<0$ and we get the final case $\Delta=0$ as 
\begin{equation}
h (\tau)=\Phi_0-\frac{2}{\tau-\tau_0}, ~A=-\frac{4}{3}.
\end{equation}

\paragraph{{N}on-conservative Linear EOS.}

The next model is based on assuming an effective equation of state, $\tensor{p}{_{eff}}=w \tensor{\rho}{_{eff}}$, for some real parameter $w \in \mathbb{R}$. Since the system involves two unknowns $(H,\phi)$, but only a single dynamical equation, an additional condition is required. Alternatively, one could follow the approach of Paiva et. al. \cite{Paiva_2022} and specify a functional form for $\phi(t)$; however, adopting an effective equation of state is a common strategy in modified gravity, and we follow this method here.

Using (\ref{rhoeff}) and (\ref{peff}) gives the equation 
\begin{align}
    \dot{\phi} &= \phi^2-2\phi H-\frac{\alpha}{2} \rho -3w\left(\phi H +\frac{\alpha }{2}\rho\right) \nonumber\\
    &= \phi^2-\frac{(32\pi+3\alpha)\phi +3\alpha H}{16\pi +3\alpha}H
    -3w\frac{16\pi \phi+3\alpha H}{16\pi +3\alpha}H \label{EOSeq1},
\end{align}
which can easily be made dimensionless via (\ref{dimlessparams}). Then, combining the dimensionless form of (\ref{EOSeq1}) with (\ref{Model1dim}) and defining a redshift parameter $z$ by $1+z=1/a$, gives the system
\begin{align} \label{Model1p}
    \frac{d\Phi(z)}{dz} &= \frac{1}{1+z}\bigg(-\frac{\Phi^2(z)}{h(z)}+3w\frac{\Phi(z)+Ah(z)}{1+A}\nonumber\\
    &+\frac{2+A}{1+A}\Phi(z)+\frac{A}{1+A}h(z)\bigg) ,
\end{align}
\begin{align}\label{Model1h}
 \frac{dh(z)}{dz} &= \frac{3}{2+2z}\left(h(z)+w\frac{\Phi(z)+Ah(z)}{1+A}\right) ,
\end{align}
where we have made use of the relation
\begin{equation}
    \frac{d}{d\tau}=-(1+z)h(z)\frac{d}{dz} .
\end{equation}

Let us note that in case of vanishing Herglotz field, this system reduces to the single equation
\begin{equation}\label{TModelnoH}
\frac{dh(z)}{dz} = \frac{B}{1+z}h(z) ,
\end{equation}
where $B = (3\alpha+24\pi)/(3\alpha+16\pi)$. This corresponds to the standard $f(R,T)=R+\alpha T$ theory that has been explored extensively, for example in \cite{Harko_2011,PhysRevD.95.123536}. The solution to Eq. (\ref{TModelnoH}) is given by 
\begin{equation}
h(z) = (1+z)^B ,
\end{equation}
which admits the constant deceleration parameter $q=B-1$. This is precisely why this model has traditionally been regarded as inadequate for describing the evolution of the Universe \cite{PhysRevD.95.123536}. However, in the presence of a non-vanishing Herglotz field, the deceleration parameter is no longer constant, as we shall see in the following.

\subsubsection{Model II: $f(R,T) = R+\alpha T^{-1}$}

The next model we consider is a specific case of $f(R,T) = R+\alpha T^\epsilon$, where we choose $\epsilon = -1$.

For this model we define $(\tau, h, \Phi,r)$ by Eq. (\ref{dimlessparams}) and let 
\begin{equation}
    \alpha = 16\pi A\left(\frac{3H_0^2}{8\pi}\right)^{2}  .
\end{equation}
Then, using Eqs. (\ref{Fr1}), (\ref{Fr2}), and a linear effective equation of state we arrive at 
\begin{eqnarray}
    h^2(\tau) &=& \Phi(\tau) h(\tau) +r(\tau) -\frac{ A}{r(\tau)}\label{epsilondensityeq} ,\\
    2\frac{dh(\tau)}{d\tau} -\frac{d\Phi(\tau)}{d\tau} &=& 2\Phi(\tau) h(\tau) -3h^2(\tau) \nonumber\\
    &&-\Phi^2(\tau)+ \frac{3A}{r(\tau)} ,\\
    \frac{d\Phi(\tau)}{d\tau} &=& \Phi^2(\tau) -2\Phi(\tau) h(\tau)-\frac{3A}{r(\tau)} \nonumber\\
    &&-3w\left(\Phi(\tau) h(\tau) -\frac{A}{r(\tau)}\right) .
\end{eqnarray}

Looking at Eq{.} (\ref{epsilondensityeq}) it is clear that an algebraic constraint on the density function will have two solutions given by 
\begin{eqnarray} \label{Constraintmodel2}
r_{\pm}(\tau) &=& \frac{1}{2} \bigg(h(\tau)\left(h(\tau)-\Phi(\tau)\right) \nonumber\\
&&\pm \sqrt{h^2\left(\tau)(h(\tau)-\Phi(\tau)\right)^2+4A}\bigg) .
\end{eqnarray}
Considering both solutions we move into a redshift representation so that the dynamical equations read
\begin{eqnarray} 
    \frac{dh}{dz} &=& \frac{3}{2+2z}\left[h+\frac{w}{2}\left(h+\Phi \mp \frac{\sqrt{}}{h}\right)\right] ,\label{model2h} \\
\frac{d\Phi}{dz} &=& \frac{1}{1+z}\left(-\frac{\Phi^2}{h}+5\Phi +\frac{3}{2}\left(w-1\right)\left(h+\Phi \mp \frac{\sqrt{}}{h}\right)\right) .\label{model2p}\nonumber \\
\end{eqnarray}
where $\sqrt{} = \sqrt{h^2(h-\Phi)^2 +4A}$.

\subsection{Numerical solutions }
Given the complexity of the models described by Eqs. (\ref{Model1p})-(\ref{Model1h}) and (\ref{model2h})-(\ref{model2p}), we cannot solve them analytically. We therefore resort to numerical techniques. The purpose of this section is to demonstrate that, for a certain range of model parameters, the considered models provide a viable description of the publicly available cosmic chronometer data \cite{moresco2020HzTable}.\footnote{We do not claim to have identified the optimal model parameters; this issue could be addressed in future work.}

We proceed as follows. For Models I and II, we integrate the systems \eqref{Model1p}-\eqref{Model1h} and \eqref{model2p}-\eqref{model2h} numerically using Mathematica, adopting initial conditions $h(0)=1$ and $\Phi(0)=\Phi_{0}$, a value to be specified later. Given the numerical solutions for $\Phi(z)$ and $h(z)$, the dimensionless matter density for Model I is obtained from Eq. \eqref{Constrainmodel1} as
\begin{equation}
    r(z)=\frac{h^2(z)-\Phi(z) h(z)}{1+A}.
\end{equation}
For Model II, the density is determined from \eqref{Constraintmodel2} by selecting the positive root of the resulting expression.

\begin{figure*}[htbp]
\centering
\includegraphics[width=0.490\linewidth]{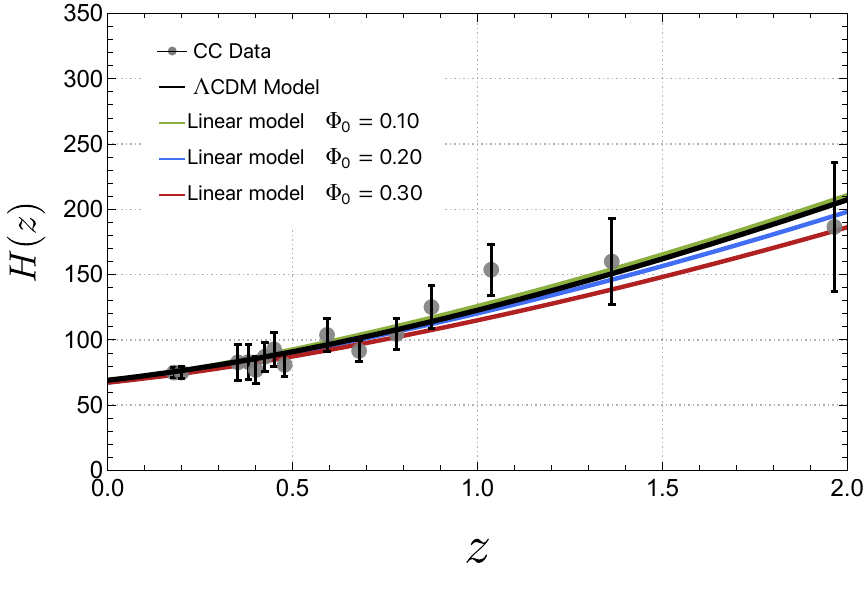} 
\includegraphics[width=0.490\linewidth]{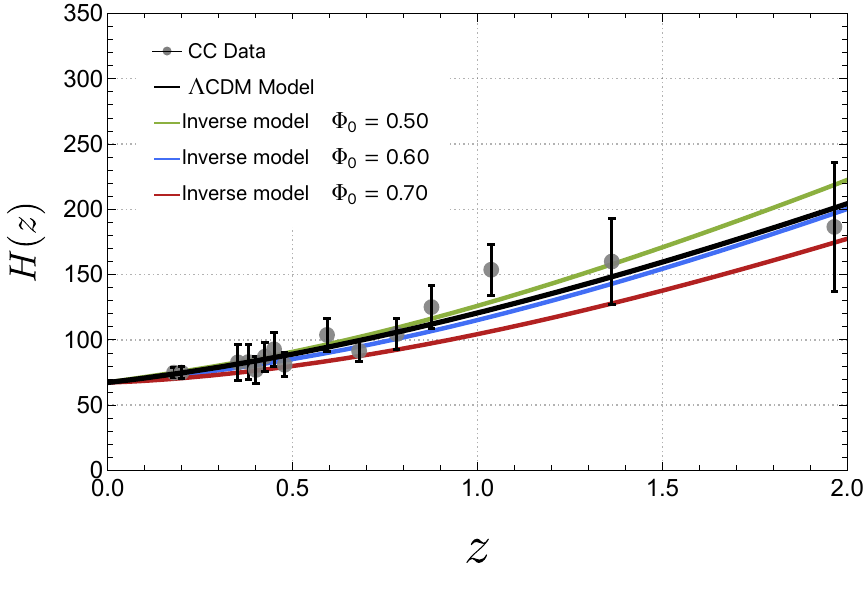}
\caption{The Hubble function $H(z)$ as a function of redshift for the models $f(R,T)=R+\alpha T$ (left panel) and $f(R,T)=R+\alpha T^{-1}$ (right panel), shown for different initial values $\tensor{\Phi}{_{0}}$.}
\label{fig1}
\end{figure*}

\begin{figure*}[htbp]
\centering
\includegraphics[width=0.490\linewidth]{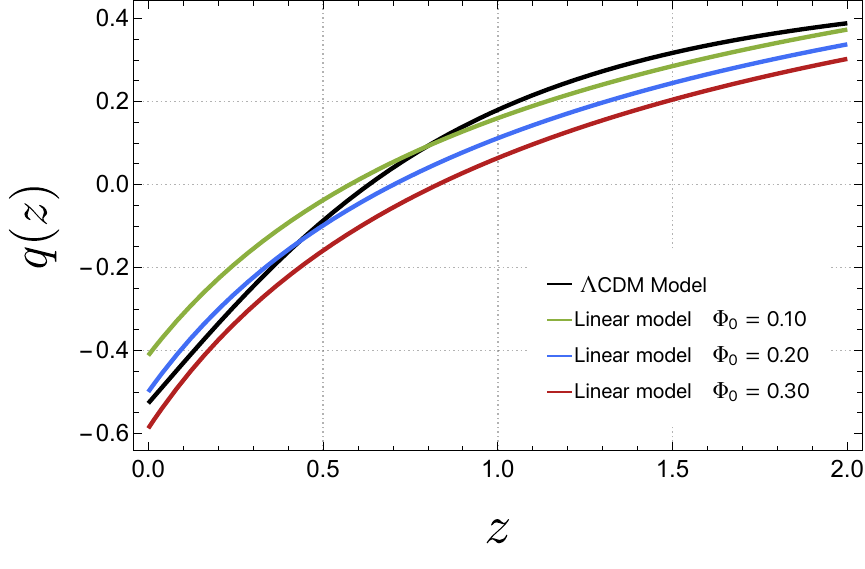} 
\includegraphics[width=0.490\linewidth]{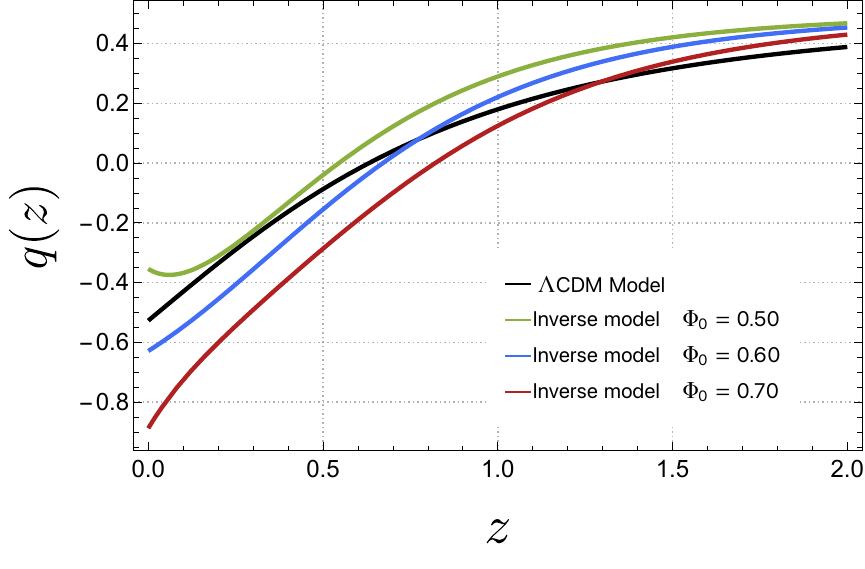}
\caption{Redshift evolution of the deceleration parameter \(q(z)\) for the models \(f(R,T)=R+\alpha T\) (left) and \(f(R,T)=R+\alpha T^{-1}\) (right), displayed for several initial values of \(\Phi_0\).}
\label{fig2}
\end{figure*}

\begin{figure*}[htbp]
\centering
\includegraphics[width=0.490\linewidth]{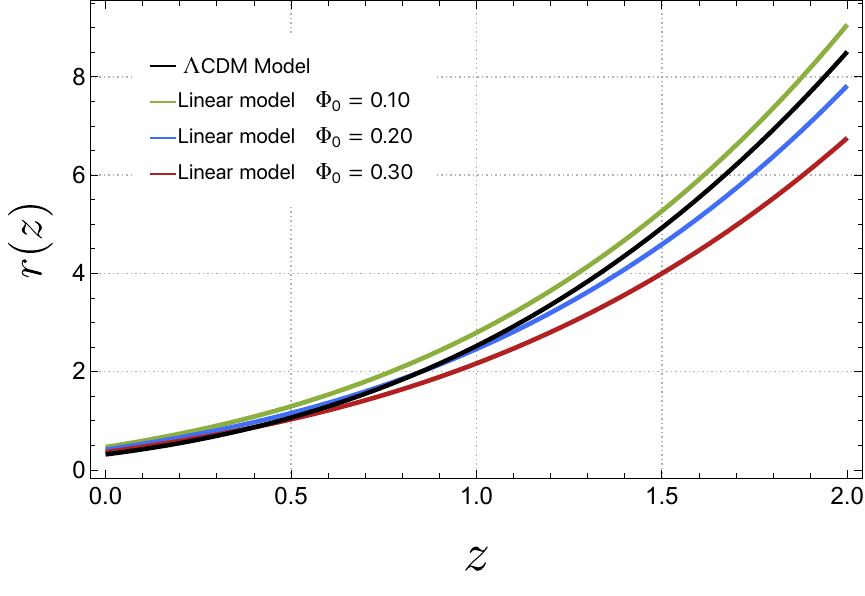} 
\includegraphics[width=0.490\linewidth]{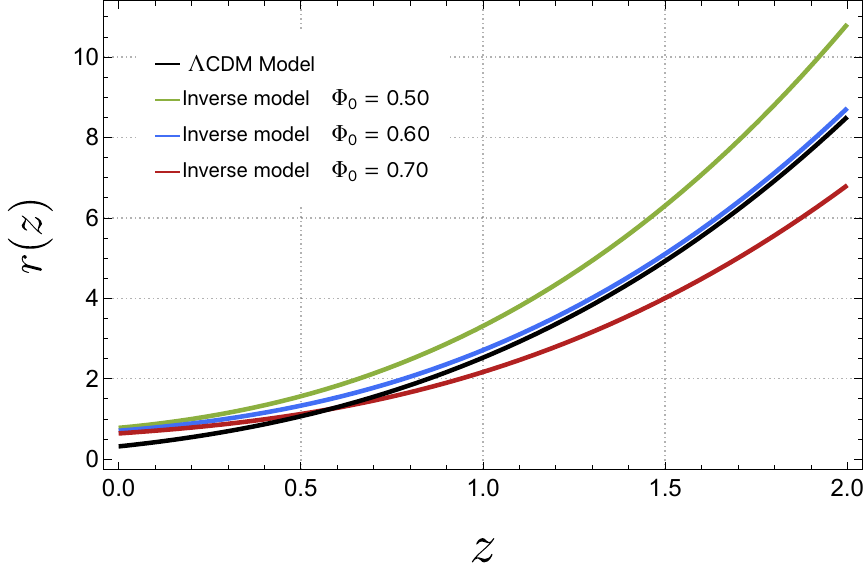}
\caption{Evolution of the dimensionless matter density \(r(z)\) with redshift for the $f(R,T)=R+\alpha T$ (left) and $f(R,T)=R+\alpha T^{-1}$ (right) models, for various initial values of \(\Phi_0\).}
\label{fig3}
\end{figure*}

\begin{figure*}[htbp]
\centering
\includegraphics[width=0.490\linewidth]{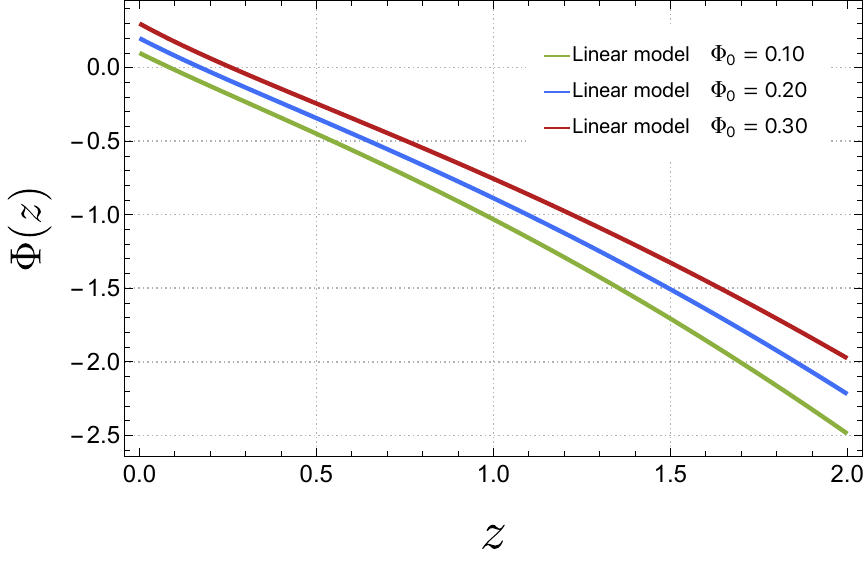} 
\includegraphics[width=0.490\linewidth]{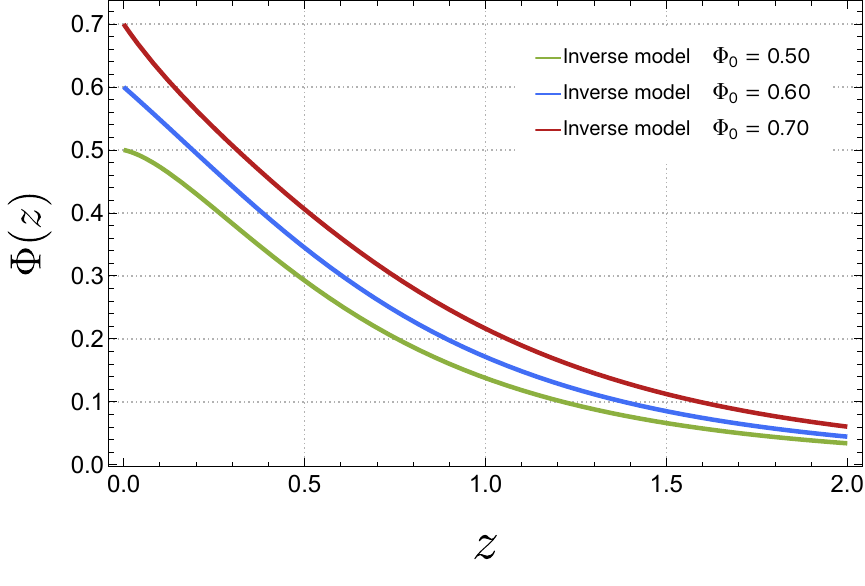}
\caption{Redshift dependence of the Herglotz contribution for the models $f(R,T)=R+\alpha T$ (left) and $f(R,T)=R+\alpha T^{-1}$ (right), plotted for different initial values of \(\Phi_0\).}
\label{fig4}
\end{figure*}

For both models, we also compute the cosmographic parameters: the deceleration parameter
\begin{equation}
    q(z)=(1+z)\frac{1}{h(z)} \frac{dh(z)}{dz}-1,
\end{equation}
the jerk parameter
\begin{equation}
    j(z)=q(z)(2q(z)+1)+(1+z)\frac{dq(z)}{dz} ,
\end{equation}
and the snap parameter
\begin{equation}
    s(z)=\frac{j(z)-1}{3\left( q(z) -\frac{1}{2} \right)} .
\end{equation}
These quantities allow us to analyze the cosmographic behavior of the models. Furthermore, to assess whether the models lie in the quintessence or phantom regimes, or permit phantom crossing, we evaluate the $Om(z)$ diagnostic
\begin{equation}
    Om(z)=\frac{h^2(z)-1}{(1+z)^3-1}.
\end{equation}
All results are compared with the standard $\Lambda$CDM model, characterized by the Hubble function
\begin{equation}
    H(z)=\tensor{H}{_0}\sqrt{\tensor{\Omega}{_m}(1+z)^3+\left(1-\tensor{\Omega}{_m}  \right)},
\end{equation}
with parameters $\tensor{H}{_0}=67.36$, $\tensor{\Omega}{_m}=0.3153$ \cite{refId0}. For the two models, we adopt the following parameter choices:
\begin{enumerate}
    \item Model I: $A=0.94, w=-1.134, H_0=67.36,\tensor{\Phi}{_{0}} \in \{0.1,0.2,0.3 \}$ ;
    \item Model II: $A=0.215, w=-2.55, H_0=67.36, \tensor{\Phi}{_{0}} \in \{0.5,0.6,0.7\}$.
\end{enumerate}

As shown in Fig.~\ref{fig1}, for low redshifts (\(z \lesssim 2\)) both models are in good agreement with the cosmic chronometer data. For specific initial values of \(\Phi_0\) (\(0.10\) for the \(f(R,T)=R+\alpha T\) model and \(0.60\) for the \(f(R,T)=R+\alpha T^{-1}\) model), the Hubble function closely matches the predictions of the \(\Lambda\)CDM model.

Fig.~\ref{fig2} shows that the deceleration parameter for the Herglotz-type $f(R,T)$ models closely follows the behavior predicted by the $\Lambda$CDM model. Notably, for the linear model $f(R,T)=R+\alpha T$ within the Herglotz framework, the deceleration parameter is no longer constant, in contrast to its behavior under the Hamiltonian variational approach.

Fig.~\ref{fig3} illustrates that the dimensionless matter densities predicted by both models closely track the $\Lambda$CDM behavior. In particular, the $f(R,T)=R+\alpha T$ model stays closer to $\Lambda$CDM at late times (\(z \simeq 0\)).

Fig.~\ref{fig4} shows the redshift evolution of the Herglotz contribution. In both models, it decreases with increasing redshift. For the $f(R,T)=R+\alpha T$ model, the Herglotz contribution remains negative throughout the cosmological evolution, whereas for the $f(R,T)=R+\alpha T^{-1}$ model it stays positive at all redshifts.

\begin{figure*}[htbp]
\centering
\includegraphics[width=0.490\linewidth]{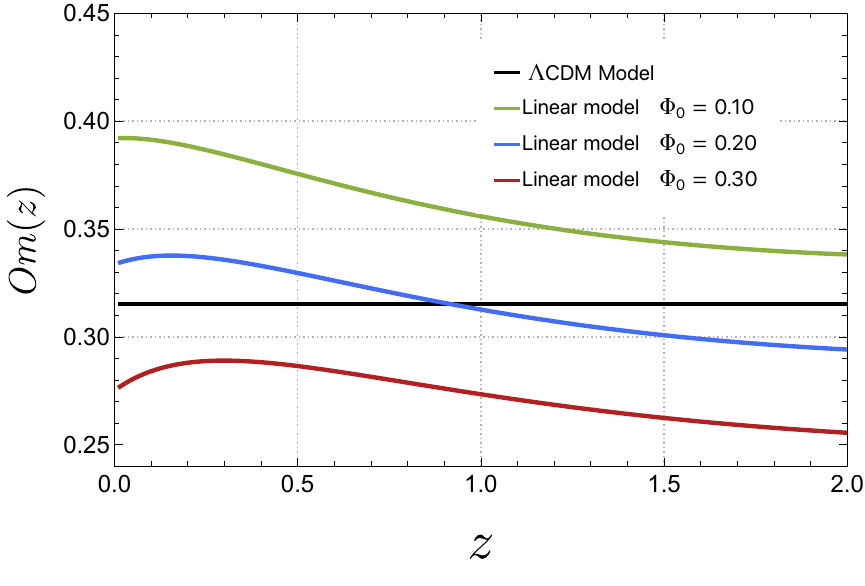} 
\includegraphics[width=0.490\linewidth]{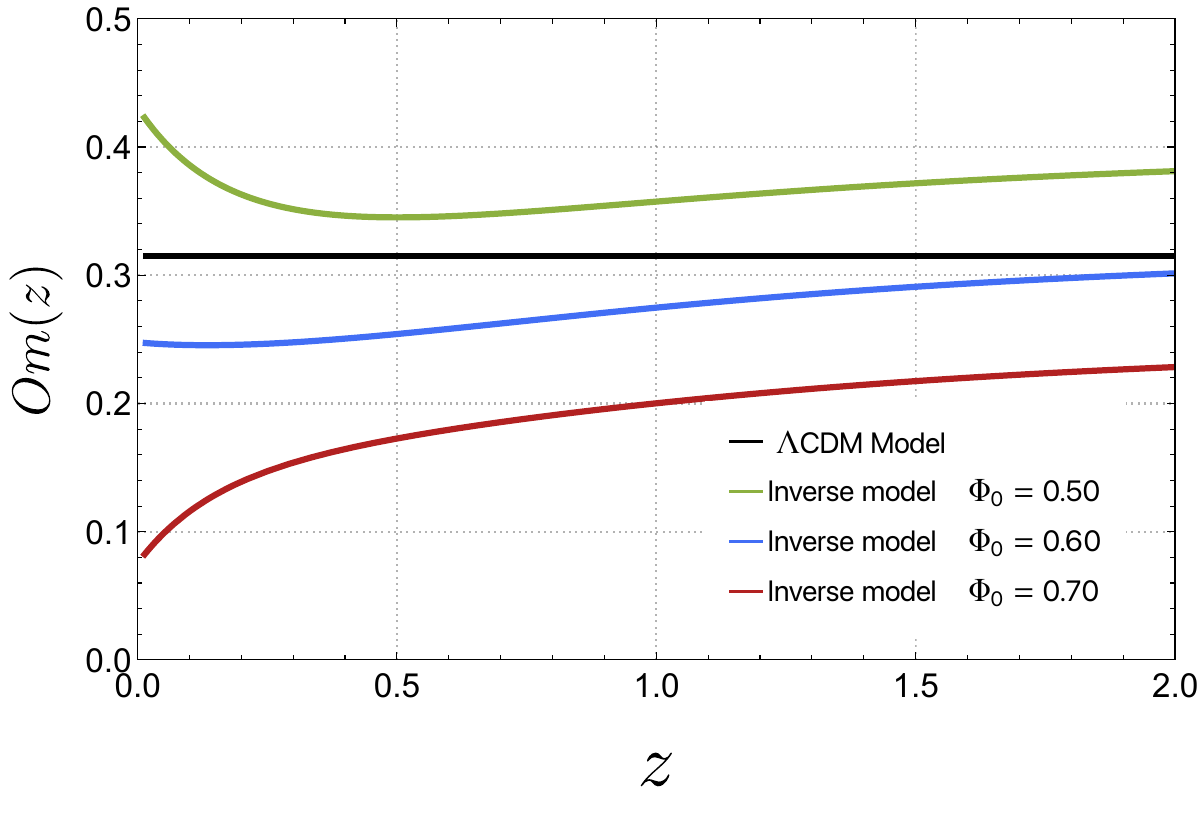}
\caption{Behavior of the $Om(z)$ diagnostic as a function of redshift for the $f(R,T)=R+\alpha T$ (left) and $f(R,T)=R+\alpha T^{-1}$ (right) models, shown for several initial values of \(\Phi_0\).}
\label{fig5}
\end{figure*}

The $Om(z)$ diagnostic is shown in Fig.~\ref{fig5}. For the $\Lambda$CDM model, it remains constant and equal to the matter density, whereas in the $f(R,T)$ models it evolves with redshift. In the $f(R,T)=R+\alpha T$ model, $Om(z)$ exhibits a positive slope at late times (typically for $z \lesssim 0.5$, depending on parameter choices), suggesting a quintessence-like behavior, and a negative slope at higher redshifts ($z \gtrsim 1$), indicative of a phantom-like regime. For the inverse model, the evolution of $Om(z)$ strongly depends on the chosen parameters; therefore, a decisive assessment requires comparison with additional datasets and strict constraints on the model parameters.

\begin{figure*}[htbp]
\centering
\includegraphics[width=0.490\linewidth]{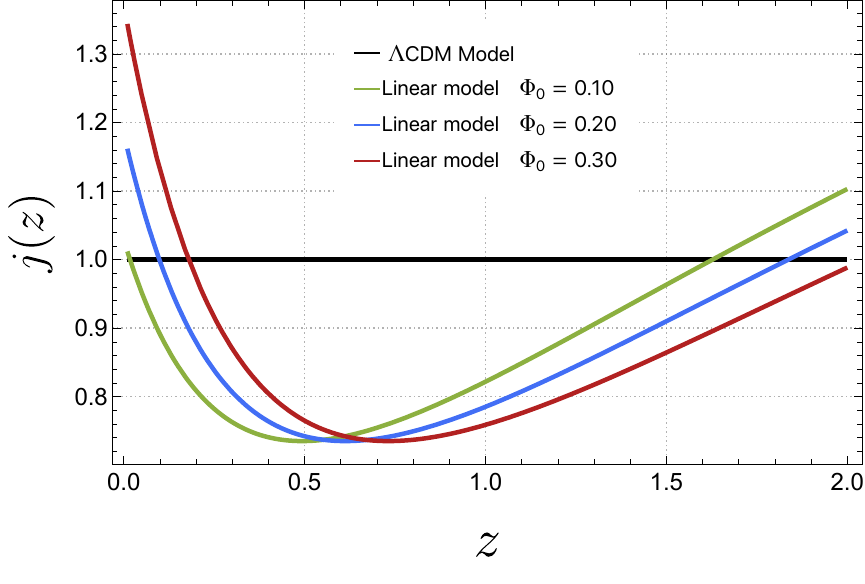} 
\includegraphics[width=0.490\linewidth]{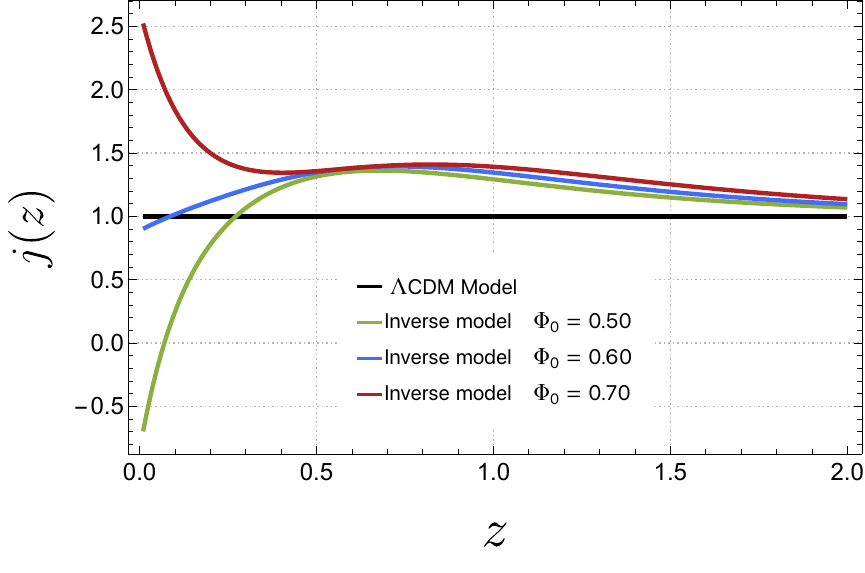}
\caption{Variation of the jerk parameter $j(z)$ for the linear cosmological model $f(R,T)=R+\alpha T$ (left panel) and for the inverse model $f(R,T)=R+\alpha T^{-1}$ (right panel), for different values of $\Phi_0$.}
\label{fig6}
\end{figure*}

Figure~\ref{fig6} shows the jerk parameter $j(z)$. In $\Lambda$CDM, $j(z) \equiv 1, \forall z \geq 0$; deviations indicate additional dynamical effects and departures from standard cosmology. In the Herglotz-type models, $j(z)$ differs from this constant, with the linear and inverse models behaving differently over redshift. For the present day, the linear model predicts $j(0) \simeq 1.18$ or $1.34$, while the inverse model gives $j(0) \simeq 2.5$ or $-0.6$, depending on the initial value of $\Phi$.

\begin{figure*}[htbp]
\centering
\includegraphics[width=0.490\linewidth]{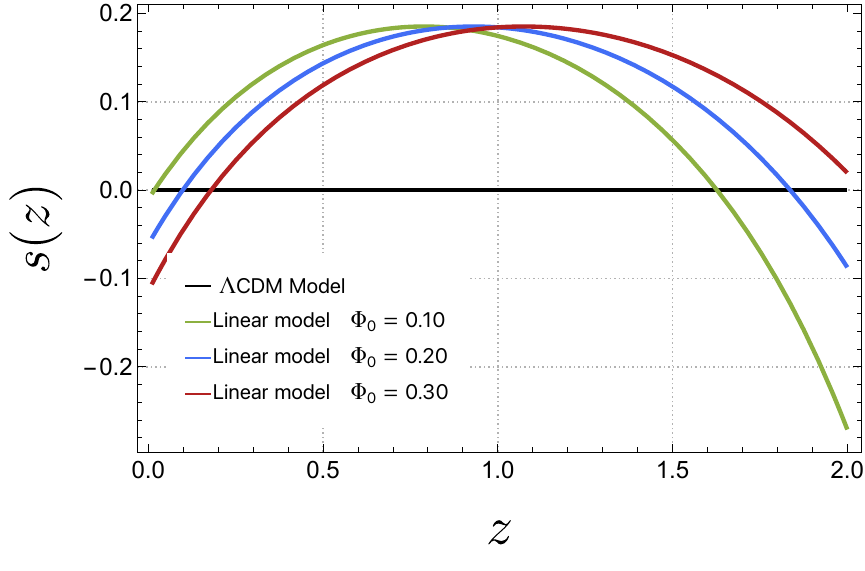} 
\includegraphics[width=0.490\linewidth]{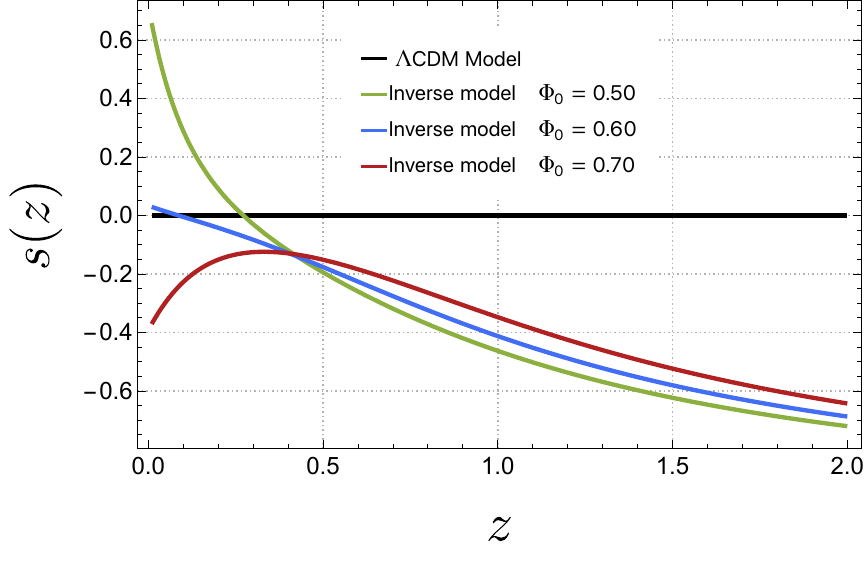}
\caption{The snap parameter $s(z)$ in terms of redshift for the linear cosmological  model $f(R,T)=R+\alpha T$ (left panel) and for the inverse cosmological model  $f(R,T)=R+\alpha T^{-1}$ (right panel), for different values of $\Phi_0$.}
\label{fig7}
\end{figure*}

The snap parameter $s(z)$, involving higher derivatives of the scale factor, provides insight into the cosmological evolution. It is particularly relevant at redshifts close to $z \simeq 2$, where expansions in powers of $z$ require terms beyond linear and quadratic. In the Herglotz-type $f(R,T)$ models, $s(z)$ differs noticeably from the $\Lambda$CDM expectation (see Fig.~\ref{fig7}). Some models show a transition from negative to positive values, while others remain negative over the entire redshift range. The present-day values are predicted as $s(0) \simeq -0.05$ or $-0.10$ for the linear model, and $s(0) \simeq -0.38$ or $0.62$ for the inverse model, with $s(0) \approx 0$ also possible. As with the jerk parameter, these values are sensitive to the initial condition $\Phi(0)=\Phi_0$.

\section{Discussion and final remarks} \label{section6}

Dissipative effects, such as viscosity and heat transfer, are essential in many physical processes. Most cosmological models, however, assume a perfect or non-dissipative fluid, an approximation that fails at high pressures or densities. Processes such as matter–radiation interactions, particle production, photon transport, and the dynamics of cosmic strings or quark–gluon plasma require dissipative dynamics. A natural extension of the matter term in the gravitational field equations is therefore the inclusion of dissipative contributions in the energy-momentum tensor, which in turn influence the gravitational field.

Quantum systems can also exhibit dissipative behavior arising from interactions with their environment \cite{CALDEIRA1983587}. A primary example is the energy decay of a thermodynamic system, often modeled as the result of irreversible energy exchange with a thermal bath. However, some decay processes occur without direct coupling to a bath and require alternative descriptions.

On the other hand one cannot a priori neglect the possibility that the gravitational field, and the associated geometry, are in fact dissipative systems, whose behavior are described by the second law of thermodynamics. Such a situation may be possible if one assumes that gravity is derived from an entropic action that couples matter fields and geometry \cite{PhysRevD.111.066001}. The entropic action can be defined as the quantum relative entropy between the metric of spacetime and the metric induced by the matter fields. 
 
Although dissipation is well known in nature, providing a Lagrangian description of dissipative phenomena presents important mathematical and physical challenges. In particular, one must distinguish between two types of Lagrangians: physical (standard) and mathematical (non-standard). A physical Lagrangian is expressed as the difference between kinetic and potential energy, while a mathematical Lagrangian reproduces the correct equations of motion without having this standard form.  

As a simple example, consider a damped oscillator describing a particle of mass $m$ moving in an external potential $U(x)$ under friction. The equation of motion Eq. \eqref{particleLeom} can also be derived from the physical Lagrangian 
\begin{equation}\label{L1}
L = e^{\nu t} \left( \frac{1}{2} m \dot{x}^2 - U(x) \right) .
\end{equation}

On the other hand the equation $\ddot{x}+\nu\dot{x}=0,$ describing the motion of a free particle in the presence of friction only,  can be obtained from both the physical (standard) Lagrangian 
\begin{equation}
L=\frac{1}{2}e^{\nu t}\dot{x}^{2}, 
\end{equation}
and from the non-physical Lagrangians  \cite{Cieslinski_2010}
\begin{equation}
L=\frac{1}{\left( e^{2\nu t}\dot{x}
+e^{\nu t}\right)}, L=\dot{x}\ln \left\vert \dot{x}\right\vert -\nu x, 
L=\left( \dot{x}^{k }+e^{-k \nu t}\right) ^{1/k },
\end{equation}
for some constant $k \in \mathbb{R}^{*}$.

Hence, there are several ways to formulate a Lagrangian description of dissipative systems. In this work, we adopt a variational approach based on the Herglotz generalization of the standard Euler-Lagrange principle. Although the Herglotz Lagrangian could in principle depend nonlinearly on the action, it remains a (generalized) physical Lagrangian, typically expressed as the difference between a kinetic term and a generalized potential. For example, the Herglotz Lagrangian for a damped particle, Eq. (\ref{particleL}) yields the same equation of motion as Eq. (\ref{L1}) and can be written as $L = T - U_{\rm eff}(x,S)$, with $U_{\rm eff}(x,S) = U(x) + (\nu/m) S$. The existence of a physical Lagrangian greatly simplifies the interpretation of the corresponding models.

In this work, we have reviewed the Herglotz variational principle in classical mechanics and  various field theories, including general relativity, and applied it to a generalized $f(R,T)$-type gravity action. This yields field equations similar to standard $f(R,T)$ gravity, but with additional contributions from the Herglotz vector. The divergence of the energy-momentum tensor was derived and found to be nonzero, as is typical in geometry–matter coupling models. However, energy-momentum conservation can be restored through a natural constraint on the Herglotz field, providing a unique mechanism to maintain generality while exploring conserved geometry–matter couplings. A natural next step would be to extend other $f(R,\text{Matter})$ \cite{Nojiri:2007bt,Mohseni:2009ns,Harko:2018gxr,Wu:2018idg,Haghani:2021fpx,deLimaJunior:2024icz} theories using the Herglotz variational approach and assess whether it yields similar improvements.

The Newtonian limit of the theory was also explored by deriving the modified Poisson equation and analyzing the motion of massive particles. In the former, corrections to the Newtonian potential were obtained for the specific cases of a constant and radial inverse square Herglotz field. The magnitude of the Herglotz field in these two cases was then constrained by considering the perihelion shift of Mercury. Additionally, the deflection angle of light around a massive body was found to scale as the square of the wavelength. In fact, this matches the scaling described by light passing through plasma which has been observed recently. Therefore, one could interpret the Herglotz field as a possible alternative description to wavelength dependent deflection near massive bodies. 

We analyzed the cosmology of Herglotz-type $f(R,T)$ gravity for a perfect fluid in a dust FLRW Universe, where the Herglotz vector reduces to a time-dependent scalar function. The generalized FLRW equations were derived, and to close the system, two forms of $f(R,T)$ were considered. For $f(R,T) = R + \alpha T$, two analytical solutions were found, corresponding to a de Sitter-like universe and a constant Herglotz function. An effective equation of state was introduced for comparison with the non-Herglotz case. Numerical solutions, with appropriately chosen parameter values, were shown to fit {a limited set of} Hubble data while producing an accelerating universe, a feature absent in the non-Herglotz scenario. The model $f(R,T) = R + \alpha T^{-1}$ was also studied using a linear effective equation of state, yielding a more complex system that required a numerical treatment.

In conclusion, the Herglotz variational extension of $f(R,T)$ gravity provides a promising framework for describing dissipative effects in gravitational systems across different scales. It also offers a good match to observational cosmological data, being able to reproduce the predictions of the $\Lambda$CDM model. However, before one could give a fair assessment of the physical and cosmological viability of Herglotz-type $f(R,T)$ models, a detailed comparison with {many} existing datasets is necessary.

From a theoretical perspective, {several foundational issues are worth investigating in the future. An important point concerns the interpretation of the Herglotz one-form $\lambda_{\mu}$. Although its presence introduces a background structure, it differs fundamentally from the vector field in Einstein-Aether theory. In Einstein-Aether models, the vector field is dynamical and varied independently within the standard variational principle. In contrast, in the Hergltoz variational approach, $\lambda_{\mu}$ arises from the covariant generalization of the variational principle itself and does not have dynamics. Nevertheless, $\lambda_{\mu}$ is not \textit{completely} arbitrary, since, by construction, it can locally be written as $\lambda_{\mu}=\partial_{\mu} \chi$ for some scalar field $\chi$. We emphasize that the non-dynamical nature of $\lambda_{\mu}$ is not specific to our model, but is inherent to the Herglotz variational framework itself. It is also important to point out that this does not restrict the applicability of the formalism to realistic physical systems, since in many dissipative processes in cosmology, dissipation is non-dynamical. For example, in the simplest formulation, bulk viscosity effects are considered in cosmology by including in the standard thermodynamic pressure $p$ a term of the form $\-\xi(t)H$, where $\xi(t)$ is the bulk viscosity coefficient, and $H$ is the Hubble function, so that an effective pressure $p_{eff}=p - \xi(t)H$ appears in the Friedmann equations. This approach corresponds to  first order or non-causal thermodynamics. In the causal thermodynamical approach, it is assumed that the effective pressure has a general form $p_{eff}=p+\Pi$, where the bulk viscous pressure satisfies a specific evolution equation, and the bulk viscosity coefficient is generically a function of density $\xi=\xi(\rho)$. Here, the deviations from equilibrium of the basic thermodynamic quantities describing dissipation (bulk stress, heat flow and shear stress) are considered as independent dynamical variables. It is also important to mention that a "viscous type" modification does appear in the Herglotz version of the generalized Friedmann equations \eqref{Frn}, and thus the Herglotz effects could be interpreted as describing generalized viscous processes, for example in the framework of causal thermodynamics with heat transfer \cite{N4}. The description of particle creation processes can also be described in the Herglotz framework, without the dynamical behaviour of the Herglotz field specified a priori. Moreover, not fixing rigidly the Herglotz field via an equation of motion gives more freedom for comparing the present formalism with various dissipative theories in which the dissipative terms are free physical parameters, not restricted by specific dynamical evolution equations, like is the case, for example, with the bulk viscosity coefficient.}

{
As pointed out in \cite{universe7020038}, since matter is not coupled to $s^{\mu}$, one should expect the Herglotz contribution to be purely geometric. We do not rule out, however, that once matter is coupled to $s^{\mu}$, one could motivate explicit choices of $\lambda_{\mu}$ from an effective field theory perspective, for example after integrating out microsopic degrees of freedom. In this sense, the present formulation remains an effective description, analogous to phenomenological approaches where dissipative effects are modeled via effective fluids or coarse-grained open system dynamics. The difference is that here dissipation is incorporated through a mathematically well-defined covariant variational principle.}

{Another foundational issue is the definition of the energy-momentum tensor in the Herglotz framework. While in this paper we adopted the standard definition to compare with previous works on $f(R,T)$ gravity and nonconservative gravity, several other approaches could be taken. An appealing approach to include the dissipative effects in the energy-momentum tensor would be to redefine $T'_{\mu\nu}:=\frac{\delta \left(\sqrt{-g} e^{-\varphi} \mathcal{L}_{m} \right)}{\delta g^{\mu \nu}}$. However, if $\varphi$ does not depend on the metric, as in the case of this paper, we simply obtain $T'_{\mu \nu}=e^{-\varphi} T_{\mu \nu}$. On the other hand, if $\varphi=\varphi\left(g_{\mu \nu} \right)$ depends on the metric, equations \eqref{GR} and \eqref{eq:variations_F_R_T} do not hold, and hence the whole theory would require a reformulation.}

{Apart from the foundational aspects, several directions for future work are natural. For example, obtaining analytic solutions in spherical symmetry, as well as exploring gravitational waves in Herglotz-type $f(R,\text{Matter})$ theories and formulating scalar-tensor representations thereof could be of interest.}

Taking into account the results of this work, Herglotz-type $f(R,T)$ gravity and the corresponding cosmological models could provide a compelling theoretical alternative to the standard $\Lambda$CDM model and general relativity. Although further work is required to clarify some of its theoretical foundations, the theory shows potential in explaining observational data and improving on models that were unrealistic under the standard Hamiltonian approach.

\section*{Acknowledgements}
MASP acknowledges support from the FCT through the Fellowship UI/BD/154479/2022 and through the project with reference PTDC/FIS-AST/0054/2021 (``BEYond LAmbda''). MASP also acknowledges support from the MICINU through the project with reference PID2023-149560NB-C21. L.Cs. and T.H. are grateful for the support of Collegium Talentum, Hungary. L.Cs. is also thankful to  Géza Tamás Szőllősi for valuable discussions.

\bibliographystyle{unsrt}
\bibliography{notes}

@article{Brading:2005ina,
    author = "Brading, Katherine",
    title = "{A Note on General Relativity, Energy Conservation, and Noether{\textquoteright}s Theorems}",
    doi = "10.1007/0-8176-4454-7_8",
    journal = "Einstein Stud.",
    volume = "11",
    pages = "125--135",
    year = "2005"
}

@book{Landau:1975pou,
    author = "Landau, L. D. and Lifschits, E. M.",
    title = "{The Classical Theory of Fields}",
    isbn = "978-0-08-018176-9",
    publisher = "Pergamon Press",
    address = "Oxford",
    series = "Course of Theoretical Physics",
    volume = "Volume 2",
    year = "1975"
}

@misc{BarroseSa:2025uxe,
      title={Clarifying the relation between covariantly conserved currents and Noether's second theorem}, 
      author={Barros e S{\'a}, Nuno and Pinto, Miguel A. S. and Trindade, Tom{\'a}s},
      year={2025},
      eprint={2506.14454},
      archivePrefix={arXiv},
      primaryClass={gr-qc},
      url={https://arxiv.org/abs/2506.14454}, 
}

@article{PhysRevD.69.064005,
  title = {Static post-Newtonian equivalence of general relativity and gravity with a dynamical preferred frame},
  author = {Eling, Christopher and Jacobson, Ted},
  journal = {Phys. Rev. D},
  volume = {69},
  issue = {6},
  pages = {064005},
  numpages = {6},
  year = {2004},
  month = {Mar},
  publisher = {American Physical Society},
  doi = {10.1103/PhysRevD.69.064005},
  url = {https://link.aps.org/doi/10.1103/PhysRevD.69.064005}
}

@article{PhysRevD.70.024003,
  title = {Einstein-aether waves},
  author = {Jacobson, T. and Mattingly, D.},
  journal = {Phys. Rev. D},
  volume = {70},
  issue = {2},
  pages = {024003},
  numpages = {5},
  year = {2004},
  month = {Jul},
  publisher = {American Physical Society},
  doi = {10.1103/PhysRevD.70.024003},
  url = {https://link.aps.org/doi/10.1103/PhysRevD.70.024003}
}

@article{Eling:2006ec,
    author = "Eling, Christopher and Jacobson, Ted",
    title = "{Black Holes in Einstein-Aether Theory}",
    doi = "10.1088/0264-9381/23/18/009",
    journal = "Class. Quant. Grav.",
    volume = "23",
    pages = "5643--5660",
    year = "2006",
    note = "[Erratum: Class.Quant.Grav. 27, 049802 (2010)]"
}

@article{Eling:2006df,
    author = "Eling, Christopher and Jacobson, Ted",
    title = "{Spherical solutions in Einstein-aether theory: Static aether and stars}",
    doi = "10.1088/0264-9381/23/18/008",
    journal = "Class. Quant. Grav.",
    volume = "23",
    pages = "5625--5642",
    year = "2006",
    note = "[Erratum: Class.Quant.Grav. 27, 049801 (2010)]"
}

@inproceedings{Eling:2004dk,
    author = "Eling, Christopher and Jacobson, Ted and Mattingly, David",
    title = "{Einstein-Aether theory}",
    booktitle = "{Deserfest: A Celebration of the Life and Works of Stanley Deser}",
    pages = "163--179",
    month = "10",
    year = "2004"
}

@article{Koivisto:2005yk,
    author = "Koivisto, Tomi",
    title = "{Covariant conservation of energy momentum in modified gravities}",
    doi = "10.1088/0264-9381/23/12/N01",
    journal = "Class. Quant. Grav.",
    volume = "23",
    pages = "4289--4296",
    year = "2006"
}

@article{Paiva_2022,
  title = {Generalized nonconservative gravitational field equations from Herglotz action principle},
  author = {Paiva, Juilson A. P. and Lazo, Matheus J. and Zanchin, Vilson T.},
  journal = {Phys. Rev. D},
  volume = {105},
  issue = {12},
  pages = {124023},
  numpages = {16},
  year = {2022},
  month = {Jun},
  publisher = {American Physical Society},
  doi = {10.1103/PhysRevD.105.124023},
  url = {https://link.aps.org/doi/10.1103/PhysRevD.105.124023}
}

@article{Harko_2011,
  title = {$f(R,T)$ gravity},
  author = {Harko, Tiberiu and Lobo, Francisco S. N. and Nojiri, Shin'ichi and Odintsov, Sergei D.},
  journal = {Phys. Rev. D},
  volume = {84},
  issue = {2},
  pages = {024020},
  numpages = {11},
  year = {2011},
  month = {Jul},
  publisher = {American Physical Society},
  doi = {10.1103/PhysRevD.84.024020},
  url = {https://link.aps.org/doi/10.1103/PhysRevD.84.024020}
}

@misc{farias2021,
      title={Cosmography of the f(R,T) gravity theory}, 
      author={I. S. Farias and P. H. R. S. Moraes},
      year={2021},
      eprint={2108.09332},
      archivePrefix={arXiv},
      primaryClass={gr-qc},
      url={https://arxiv.org/abs/2108.09332}, 
}

@article{Bouali_2023,
    author = {Bouali, Amine and Chaudhary, Himanshu and Harko, Tiberiu and Lobo, Francisco S N and Ouali, Taoufik and Pinto, Miguel A S},
    title = {Observational constraints and cosmological implications of scalar–tensor f(R, T) gravity},
    journal = {Monthly Notices of the Royal Astronomical Society},
    volume = {526},
    number = {3},
    pages = {4192-4208},
    year = {2023},
    month = {10},
    issn = {0035-8711},
    doi = {10.1093/mnras/stad2998},
    url = {https://doi.org/10.1093/mnras/stad2998},
    eprint = {https://academic.oup.com/mnras/article-pdf/526/3/4192/52191634/stad2998.pdf},
}

@article{Lazo_2018,
    author = {Lazo, Matheus J. and Paiva, Juilson and Amaral, João T. S. and Frederico, Gastão S. F.},
    title = {An action principle for action-dependent Lagrangians: Toward an action principle to non-conservative systems},
    journal = {Journal of Mathematical Physics},
    volume = {59},
    number = {3},
    pages = {032902},
    year = {2018},
    month = {03},
    issn = {0022-2488},
    doi = {10.1063/1.5019936},
    url = {https://doi.org/10.1063/1.5019936},
    eprint = {https://pubs.aip.org/aip/jmp/article-pdf/doi/10.1063/1.5019936/14039324/032902\_1\_online.pdf},
}

@ARTICLE{Bauer1931-ek,
  title    = "Dissipative Dynamical Systems: {I}",
  author   = "Bauer, P S",
  journal  = "Proc Natl Acad Sci U S A",
  volume   =  17,
  number   =  5,
  pages    = "311--314",
  month    =  may,
  year     =  1931,
  address  = "United States",
  language = "en"
}

@article{Lazo_2017,
  title = {Action principle for action-dependent Lagrangians toward nonconservative gravity: Accelerating universe without dark energy},
  author = {Lazo, Matheus J. and Paiva, Juilson and Amaral, Jo\~ao T. S. and Frederico, Gast\~ao S. F.},
  journal = {Phys. Rev. D},
  volume = {95},
  issue = {10},
  pages = {101501},
  numpages = {5},
  year = {2017},
  month = {May},
  publisher = {American Physical Society},
  doi = {10.1103/PhysRevD.95.101501},
  url = {https://link.aps.org/doi/10.1103/PhysRevD.95.101501}
}

@Article{Singh2016,
author={Singh, Vijay
and Singh, C. P.},
title={Friedmann Cosmology with Matter Creation in Modified f(R, T) Gravity},
journal={International Journal of Theoretical Physics},
year={2016},
month={Feb},
day={01},
volume={55},
number={2},
pages={1257-1273},
issn={1572-9575},
doi={10.1007/s10773-015-2767-z},
url={https://doi.org/10.1007/s10773-015-2767-z}
}

@Article{Singh2015,
author={Singh, Vijay
and Singh, C. P.},
title={Modified f(R,T) gravity theory and scalar field cosmology},
journal={Astrophysics and Space Science},
year={2015},
month={Mar},
day={01},
volume={356},
number={1},
pages={153-162},
issn={1572-946X},
doi={10.1007/s10509-014-2183-5},
url={https://doi.org/10.1007/s10509-014-2183-5}
}

@article{Harko_2014,
  title = {Thermodynamic interpretation of the generalized gravity models with geometry-matter coupling},
  author = {Harko, Tiberiu},
  journal = {Phys. Rev. D},
  volume = {90},
  issue = {4},
  pages = {044067},
  numpages = {13},
  year = {2014},
  month = {Aug},
  publisher = {American Physical Society},
  doi = {10.1103/PhysRevD.90.044067},
  url = {https://link.aps.org/doi/10.1103/PhysRevD.90.044067}
}

@article{Asadiyan_2019,
author = {Asadiyan, K. and Darabi, F. and Pedram, P. and Atazadeh, K.},
title = {Bounding $f(R,T)$ gravity by particle creation},
journal = {International Journal of Geometric Methods in Modern Physics},
volume = {16},
number = {10},
pages = {1950153},
year = {2019},
doi = {10.1142/S0219887819501536},

URL = { 
    
        https://doi.org/10.1142/S0219887819501536
    
    

},
eprint = { 
    
        https://doi.org/10.1142/S0219887819501536
    
    

}
}

@article{dosSantos2018,
    author = "dos Santos, S. I. and Carvalho, G. A. and Moraes, P. H. R. S. and Lenzi, C. H. and Malheiro, M.",
    title = "{A conservative energy-momentum tensor in the $f(R,T)$ gravity and its implications for the phenomenology of neutron stars}",
    eprint = "1803.07719",
    archivePrefix = "arXiv",
    primaryClass = "gr-qc",
    doi = "10.1140/epjp/i2019-12830-8",
    journal = "Eur. Phys. J. Plus",
    volume = "134",
    number = "8",
    pages = "398",
    year = "2019"
}

@article{Carvalho2019,
    author = "Carvalho, G. A. and Dos Santos, S. I. and Moraes, P. H. R. S. and Malheiro, M.",
    title = "{Strange stars in energy-momentum-conserved $f(R,T)$ gravity}",
    eprint = "1911.02484",
    archivePrefix = "arXiv",
    primaryClass = "gr-qc",
    doi = "10.1142/S0218271820500753",
    journal = "Int. J. Mod. Phys. D",
    volume = "29",
    number = "10",
    pages = "2050075",
    year = "2020"
}

@article{Alvarenga_2013,
  title = {Dynamics of scalar perturbations in $f(R,T)$ gravity},
  author = {Alvarenga, F. G. and de la Cruz-Dombriz, A. and Houndjo, M. J. S. and Rodrigues, M. E. and S\'aez-G\'omez, D.},
  journal = {Phys. Rev. D},
  volume = {87},
  issue = {10},
  pages = {103526},
  numpages = {9},
  year = {2013},
  month = {May},
  publisher = {American Physical Society},
  doi = {10.1103/PhysRevD.87.103526},
  url = {https://link.aps.org/doi/10.1103/PhysRevD.87.103526}
}

@Article{Chakraborty2013,
author={Chakraborty, Subenoy},
title={An alternative f(R, T) gravity theory and the dark energy problem},
journal={General Relativity and Gravitation},
year={2013},
month={Oct},
day={01},
volume={45},
number={10},
pages={2039-2052},
issn={1572-9532},
doi={10.1007/s10714-013-1577-y},
url={https://doi.org/10.1007/s10714-013-1577-y}
}

@article{Pinto_2022,
  title = {Gravitationally induced particle production in scalar-tensor $f(R,T)$ gravity},
  author = {Pinto, Miguel A. S. and Harko, Tiberiu and Lobo, Francisco S. N.},
  journal = {Phys. Rev. D},
  volume = {106},
  issue = {4},
  pages = {044043},
  numpages = {16},
  year = {2022},
  month = {Aug},
  publisher = {American Physical Society},
  doi = {10.1103/PhysRevD.106.044043},
  url = {https://link.aps.org/doi/10.1103/PhysRevD.106.044043}
}

@article{Siggia_2025,
  title = {Comparison of $f(R,T)$ gravity with type Ia supernovae data},
  author = {Siggia, Vincent R. and Carlson, Eric D.},
  journal = {Phys. Rev. D},
  volume = {111},
  issue = {2},
  pages = {024074},
  numpages = {7},
  year = {2025},
  month = {Jan},
  publisher = {American Physical Society},
  doi = {10.1103/PhysRevD.111.024074},
  url = {https://link.aps.org/doi/10.1103/PhysRevD.111.024074}
}

@article{PhysRevD.88.044023,
  title = {Further matters in space-time geometry: $f(R,T,{R}_{\ensuremath{\mu}\ensuremath{\nu}}{T}^{\ensuremath{\mu}\ensuremath{\nu}})$ gravity},
  author = {Haghani, Zahra and Harko, Tiberiu and Lobo, Francisco S. N. and Sepangi, Hamid Reza and Shahidi, Shahab},
  journal = {Phys. Rev. D},
  volume = {88},
  issue = {4},
  pages = {044023},
  numpages = {18},
  year = {2013},
  month = {Aug},
  publisher = {American Physical Society},
  doi = {10.1103/PhysRevD.88.044023},
  url = {https://link.aps.org/doi/10.1103/PhysRevD.88.044023}
}

@article{Harko:2010mv,
    author = "Harko, Tiberiu and Lobo, Francisco S. N.",
    title = "{f(R,$L_{m}$) gravity}",
    doi = "10.1140/epjc/s10052-010-1467-3",
    journal = "Eur. Phys. J. C",
    volume = "70",
    pages = "373--379",
    year = "2010"
}

@article{Harko:2014aja,
    author = "Harko, Tiberiu and Lobo, Francisco S. N. and Otalora, G. and Saridakis, Emmanuel N.",
    title = "{$f(T,\mathcal{T})$ gravity and cosmology}",
    doi = "10.1088/1475-7516/2014/12/021",
    journal = "JCAP",
    volume = "12",
    number=12,
    pages = "021",
    year = "2014"
}

@article{Cipriano:2024jng,
    author = "Cipriano, Ricardo A. C. and Ganiyeva, Nailya and Harko, Tiberiu and Lobo, Francisco S. N. and Pinto, Miguel A. S. and Rosa, Jo{\~a}o Lu{\'\i}s",
    title = "{Energy-Momentum Squared Gravity: A Brief Overview}",
    doi = "10.3390/universe10090339",
    journal = "Universe",
    volume = "10",
    number = "9",
    pages = "339",
    year = "2024"
}

@article{Bertolami:2017svl,
    author = "Bertolami, Orfeu and Gomes, Cl{\'a}udio and Lobo, Francisco S. N.",
    title = "{Gravitational waves in theories with a non-minimal curvature-matter coupling}",
    doi = "10.1140/epjc/s10052-018-5781-5",
    journal = "Eur. Phys. J. C",
    volume = "78",
    number = "4",
    pages = "303",
    year = "2018"
}

@article{AZIZI2023101303,
title = {Evolution of gravitational waves in non-minimal coupling between geometry and matter theories of gravity},
journal = {Physics of the Dark Universe},
volume = {42},
pages = {101303},
year = {2023},
issn = {2212-6864},
doi = {https://doi.org/10.1016/j.dark.2023.101303},
url = {https://www.sciencedirect.com/science/article/pii/S2212686423001371},
author = {Tahereh Azizi and Najibe Borhani and Mojtaba Haghshenas}
}

@article{PhysRevD.111.024014,
  title = {Gravitational wave polarizations in nonminimally coupled gravity},
  author = {Barroso Varela, Miguel and Bertolami, Orfeu},
  journal = {Phys. Rev. D},
  volume = {111},
  issue = {2},
  pages = {024014},
  numpages = {23},
  year = {2025},
  month = {Jan},
  publisher = {American Physical Society},
  doi = {10.1103/PhysRevD.111.024014},
  url = {https://link.aps.org/doi/10.1103/PhysRevD.111.024014}
}

@article{Deb:2018gzt,
    author = "Deb, Debabrata and Ketov, Sergei V. and Khlopov, Maxim and Ray, Saibal",
    title = "{Study on charged strange stars in $f(R, T)$ gravity}",
    doi = "10.1088/1475-7516/2019/10/070",
    journal = "JCAP",
    volume = "10",
    number= 10,
    pages = "070",
    year = "2019"
}

@article{Pinto:2025loq,
    author = "Pinto, Miguel A. S. and Maluf, Roberto V. and Olmo, Gonzalo J.",
    title = "{Regular black hole solutions in $(2 + 1)$-dimensional f(R,~T) gravity coupled to nonlinear electrodynamics}",
    doi = "10.1140/epjc/s10052-025-14585-0",
    journal = "Eur. Phys. J. C",
    volume = "85",
    number = "8",
    pages = "835",
    year = "2025"
}

@article{Tretyakov:2018yph,
    author = "Tretyakov, Petr V.",
    title = "{Cosmology in modified $f(R,T)$-gravity}",
    doi = "10.1140/epjc/s10052-018-6367-y",
    journal = "Eur. Phys. J. C",
    volume = "78",
    number = "11",
    pages = "896",
    year = "2018"
}

@article{PhysRevD.98.064045,
  title = {Big-bang nucleosynthesis and cosmic microwave background constraints on nonminimally coupled theories of gravity},
  author = {Azevedo, R. P. L. and Avelino, P. P.},
  journal = {Phys. Rev. D},
  volume = {98},
  issue = {6},
  pages = {064045},
  numpages = {5},
  year = {2018},
  month = {Sep},
  publisher = {American Physical Society},
  doi = {10.1103/PhysRevD.98.064045},
  url = {https://link.aps.org/doi/10.1103/PhysRevD.98.064045}
}

@article{PhysRevD.102.084051,
  title = {Nonminimally coupled Boltzmann equation: Foundations},
  author = {Bertolami, Orfeu and Gomes, Cl\'audio},
  journal = {Phys. Rev. D},
  volume = {102},
  issue = {8},
  pages = {084051},
  numpages = {7},
  year = {2020},
  month = {Oct},
  publisher = {American Physical Society},
  doi = {10.1103/PhysRevD.102.084051},
  url = {https://link.aps.org/doi/10.1103/PhysRevD.102.084051}
}

@article{Farias:2021jdz,
    author = "Farias, I. S. and Moraes, P. H. R. S.",
    title = "{Cosmography of $\boldsymbol{f(R,T)}$ Gravity}",
    doi = "10.1134/S0202289324010055",
    journal = "Grav. Cosmol.",
    volume = "30",
    number = "1",
    pages = "28--39",
    year = "2024"
}

@article{Harko:2015pma,
    author = "Harko, Tiberiu and Lobo, Francisco S. N. and Mimoso, Jos{\'e} P. and Pav{\'o}n, Diego",
    title = "{Gravitational induced particle production through a nonminimal curvature-matter coupling}",
    doi = "10.1140/epjc/s10052-015-3620-5",
    journal = "Eur. Phys. J. C",
    volume = "75",
    pages = "386",
    year = "2015"
}

@article{Rosa:2022osy,
    author = "Rosa, Jo{\~a}o Lu{\'\i}s and Kull, Paul Martin",
    title = "{Non-exotic traversable wormhole solutions in linear $f\left( R,T\right) $ gravity}",
    doi = "10.1140/epjc/s10052-022-11135-w",
    journal = "Eur. Phys. J. C",
    volume = "82",
    number = "12",
    pages = "1154",
    year = "2022"
}

@article{Rosa:2023guo,
    author = "Rosa, Jo{\~a}o Lu{\'\i}s and Ganiyeva, Nailya and Lobo, Francisco S. N.",
    title = "{Non-exotic traversable wormholes in $f\left( R,T_{ab}T^{ab}\right) $ gravity}",
    doi = "10.1140/epjc/s10052-023-12232-0",
    journal = "Eur. Phys. J. C",
    volume = "83",
    number = "11",
    pages = "1040",
    year = "2023"
}

@misc{Herg1,
  author       = {G. Herglotz},
  title        = {Ber\"uhrungstransformationen},
  howpublished = {Lectures at the University of G\"ottingen},
  address      = {G\"ottingen},
  year         = {1930}
}

@book{Herg2,
  author       = {R. B. Guenther and C. M. Guenther and J. A. Gottsch},
  title        = {Lecture Notes in Nonlinear Analysis, Vol.~1: The Herglotz Lectures on Contact Transformations and Hamiltonian Systems},
  publisher    = {Nicholas Copernicus University},
  address      = {Tor\'un},
  year         = {1996}
}

@article{Paliathanasis:2020plf,
    author = "Paliathanasis, Andronikos and Leon, Genly and Barrow, John D.",
    title = "{Einstein-aether theory in Weyl integrable geometry}",
    doi = "10.1140/epjc/s10052-020-08598-0",
    journal = "Eur. Phys. J. C",
    volume = "80",
    number = "12",
    pages = "1099",
    year = "2020"
}

@article{Paliathanasis:2021qns,
    author = "Paliathanasis, Andronikos and Leon, Genly",
    title = "{Integrability and cosmological solutions in Einstein-{\ae}ther-Weyl theory}",
    doi = "10.1140/epjc/s10052-021-09031-w",
    journal = "Eur. Phys. J. C",
    volume = "81",
    number = "3",
    pages = "255",
    year = "2021"
}

@article{Paliathanasis:2020pax,
    author = "Paliathanasis, Andronikos and Leon, Genly",
    title = "{Dynamics and exact Bianchi I spacetimes in Einstein-aether scalar field theory}",
    doi = "10.1140/epjc/s10052-020-8148-7",
    journal = "Eur. Phys. J. C",
    volume = "80",
    number = "6",
    pages = "589",
    year = "2020"
}

@article{Paliathanasis:2020axi,
    author = "Paliathanasis, Andronikos and Leon, Genly",
    title = "{Analytic solutions in Einstein-aether scalar field cosmology}",
    doi = "10.1140/epjc/s10052-020-7924-8",
    journal = "Eur. Phys. J. C",
    volume = "80",
    number = "5",
    pages = "355",
    year = "2020"
}

@article{PhysRevD.103.044001,
  title = {Exact black hole solutions in Einstein-aether scalar field theory},
  author = {Dimakis, N. and Leon, Genly and Paliathanasis, Andronikos},
  journal = {Phys. Rev. D},
  volume = {103},
  issue = {4},
  pages = {044001},
  numpages = {10},
  year = {2021},
  month = {Feb},
  publisher = {American Physical Society},
  doi = {10.1103/PhysRevD.103.044001},
  url = {https://link.aps.org/doi/10.1103/PhysRevD.103.044001}
}

@article{Leon:2019jnu,
    author = "Leon, Genly and Coley, A. and Paliathanasis, Andronikos",
    title = "{Static spherically symmetric Einstein-aether models II: Integrability and the modified Tolman-Oppenheimer-Volkoff approach}",
    doi = "10.1016/j.aop.2019.168002",
    journal = "Annals Phys.",
    volume = "412",
    pages = "168002",
    year = "2020"
}

@article{Paliathanasis:2019pcl,
    author = "Paliathanasis, Andronikos and Papagiannopoulos, G. and Basilakos, Spyros and Barrow, John D.",
    title = "{Dynamics of Einstein-Aether scalar field cosmology}",
    doi = "10.1140/epjc/s10052-019-7229-y",
    journal = "Eur. Phys. J. C",
    volume = "79",
    number = "8",
    pages = "723",
    year = "2019"
}

@article{PhysRevD.95.123536,
  title = {Cosmological inviability of $f(R,T)$ gravity},
  author = {Velten, Hermano and Caram\^es, Thiago R. P.},
  journal = {Phys. Rev. D},
  volume = {95},
  issue = {12},
  pages = {123536},
  numpages = {9},
  year = {2017},
  month = {Jun},
  publisher = {American Physical Society},
  doi = {10.1103/PhysRevD.95.123536},
  url = {https://link.aps.org/doi/10.1103/PhysRevD.95.123536}
}

@article{Fabris_2018,
author = {Fabris, J\'{u}lio C. and Velten, Hermano and Caram\^{e}s, Thiago R. P. and Lazo, Matheus J. and Frederico, Gast\~{a}o S. F.},
title = {Cosmology from a new nonconservative gravity},
journal = {International Journal of Modern Physics D},
volume = {27},
number = {06},
pages = {1841006},
year = {2018},
doi = {10.1142/S0218271818410067}
}

@article{Harko:2023tzo,
    author = "Harko, Tiberiu",
    title = "{Dissipative quintessence and its cosmological implications}",
    doi = "10.1103/PhysRevD.107.123507",
    journal = "Phys. Rev. D",
    volume = "107",
    number = "12",
    pages = "123507",
    year = "2023"
}

@article{Slava2019,
  title = {Diffraction of light by the gravitational field of the Sun and the solar corona},
  author = {Turyshev, Slava G. and Toth, Viktor T.},
  journal = {Phys. Rev. D},
  volume = {99},
  issue = {2},
  pages = {024044},
  numpages = {53},
  year = {2019},
  month = {Jan},
  publisher = {American Physical Society},
  doi = {10.1103/PhysRevD.99.024044},
  url = {https://link.aps.org/doi/10.1103/PhysRevD.99.024044}
}

@article{Clegg_1998,
doi = {10.1086/305344},
url = {https://doi.org/10.1086/305344},
year = {1998},
month = {mar},
publisher = {},
volume = {496},
number = {1},
pages = {253},
author = {Clegg, Andrew W. and Fey, Alan L. and Lazio, T. Joseph W.},
title = {The Gaussian Plasma Lens in Astrophysics: Refraction},
journal = {The Astrophysical Journal}
}

@book{Perlick2000,
  author    = {Volker Perlick},
  title     = {Ray Optics, Fermat’s Principle, and Applications to General Relativity},
  series    = {Lecture Notes in Physics Monographs},
  volume    = {61},
  publisher = {Springer-Verlag Berlin Heidelberg},
  year      = {2000},
  isbn      = {3-540-66898-5},
  doi       = {10.1007/3-540-46662-2}
}

@article{Bertotti2003,
  author    = {Bertotti, B. and Iess, L. and Tortora, P.},
  title     = {A test of general relativity using radio links with the Cassini spacecraft},
  journal   = {Nature},
  year      = {2003},
  volume    = {425},
  number    = {6956},
  pages     = {374--376},
  doi       = {10.1038/nature01997},
  url       = {https://doi.org/10.1038/nature01997},
  issn      = {1476-4687}
}

@article{ArnauMas,
title = {A variational derivation of the field equations of an action-dependent Einstein-Hilbert Lagrangian},
journal = {Journal of Geometric Mechanics},
volume = {15},
number = {1},
pages = {357-374},
year = {2023},
issn = {1941-4889},
doi = {10.3934/jgm.2023014},
url = {https://www.aimspress.com/article/doi/10.3934/jgm.2023014},
author = {Jordi Gaset and Arnau Mas},
}

@article{Brill1999,
    author = {Brill, Dieter R. and Goel, Deepak},
    title = {Light bending and perihelion precession: A unified approach},
    journal = {American Journal of Physics},
    volume = {67},
    number = {4},
    pages = {316-319},
    year = {1999},
    month = {04},
    issn = {0002-9505},
    doi = {10.1119/1.19255}
}

@book{Carroll_2019, place={Cambridge}, title={Spacetime and Geometry: An Introduction to General Relativity}, publisher={Cambridge University Press}, author={Carroll, Sean M.}, year={2019}}

@Article{universe7020038,
AUTHOR = {Velten, Hermano and Caramês, Thiago R. P.},
TITLE = {To Conserve, or Not to Conserve: A Review of Nonconservative Theories of Gravity},
JOURNAL = {Universe},
VOLUME = {7},
YEAR = {2021},
NUMBER = {2},
ARTICLE-NUMBER = {38},
ISSN = {2218-1997},
DOI = {10.3390/universe7020038}
}

@article{PhysRevD.53.5483,
  title = {Bulk viscous cosmology},
  author = {Zimdahl, Winfried},
  journal = {Phys. Rev. D},
  volume = {53},
  issue = {10},
  pages = {5483--5493},
  numpages = {0},
  year = {1996},
  month = {May},
  publisher = {American Physical Society},
  doi = {10.1103/PhysRevD.53.5483}
}

@article{FabrisBraneWorld,
  author  = {Fabris, J. C. and Caram{\^e}s, T. R. P. and da Silva, J. M. H.},
  title   = {Braneworld gravity within non-conservative gravitational theory},
  journal = {Eur. Phys. J. C},
  volume  = {78},
  pages   = {402},
  year    = {2018},
  doi     = {10.1140/epjc/s10052-018-5891-0}
}

@article{Braganca2019,
  author  = {Bragança, E. A. F. and Caram{\^e}s, T. R. P. and Fabris, J. C. and others},
  title   = {Some effects of non-conservative gravity on cosmic string configurations},
  journal = {Eur. Phys. J. C},
  volume  = {79},
  pages   = {162},
  year    = {2019},
  doi     = {10.1140/epjc/s10052-019-6672-0}
}

@article{PhysRevD.99.124031,
  title = {Existence of static spherically-symmetric objects in action-dependent Lagrangian theories},
  author = {Fabris, Julio C. and Velten, Hermano and Wojnar, Aneta},
  journal = {Phys. Rev. D},
  volume = {99},
  issue = {12},
  pages = {124031},
  numpages = {6},
  year = {2019},
  month = {Jun},
  publisher = {American Physical Society},
  doi = {10.1103/PhysRevD.99.124031}
}

@article{PhysRevD.103.044018,
  title = {Wormhole geometries induced by action-dependent Lagrangian theories},
  author = {Ayuso, Ismael and Lobo, Francisco S. N. and Mimoso, Jos\'e P.},
  journal = {Phys. Rev. D},
  volume = {103},
  issue = {4},
  pages = {044018},
  numpages = {12},
  year = {2021},
  month = {Feb},
  publisher = {American Physical Society},
  doi = {10.1103/PhysRevD.103.044018},
}

@article{PhysRevD.110.124056,
  title = {Characterization of wormhole spacetimes supported by a covariant action-dependent Lagrangian theory},
  author = {Ayuso, Ismael and Lazkoz, Ruth},
  journal = {Phys. Rev. D},
  volume = {110},
  issue = {12},
  pages = {124056},
  numpages = {15},
  year = {2024},
  month = {Dec},
  publisher = {American Physical Society},
  doi = {10.1103/PhysRevD.110.124056}
}

@article{PhysRevD.107.124005,
  title = {Fully conservative $f(R,T)$ gravity and Solar System constraints},
  author = {Bertini, Nicolas R. and Velten, Hermano},
  journal = {Phys. Rev. D},
  volume = {107},
  issue = {12},
  pages = {124005},
  numpages = {8},
  year = {2023},
  month = {Jun},
  publisher = {American Physical Society},
  doi = {10.1103/PhysRevD.107.124005}
}

@article{deLeon:2024ztn,
    author = "de Le{\'o}n, Manuel and Gaset Rif{\`a}, Jordi and Mu{\~n}oz-Lecanda, Miguel C. and Rivas, Xavier and Rom{\'a}n-Roy, Narciso",
    title = "{Practical Introduction to Action-Dependent Field Theories}",
    primaryClass = "hep-th",
    doi = "10.1002/prop.70000",
    journal = "Fortsch. Phys.",
    volume = "73",
    number = "5",
    pages = "e70000",
    year = "2025"
}

@book{French1968SpecialRelativity,
  author    = {A. P. French},
  title     = {Special Relativity},
  edition   = {1st},
  year      = {1968},
  publisher = {CRC Press},
  doi       = {10.1201/9781315272597}
}

@article{doi:10.1142/S0219887822501560,
author = {Gaset, Jordi and Mar\'{\i}n-Salvador, Adri\`{a}},
title = {Application of Herglotz’s variational principle to electromagnetic systems with dissipation},
journal = {International Journal of Geometric Methods in Modern Physics},
volume = {19},
number = {10},
pages = {2250156},
year = {2022},
doi = {10.1142/S0219887822501560}
}

@misc{moresco2020HzTable,
  author       = {Moresco, Michele},
  title        = {CCcovariance: Data - HzTable\_MM\_BC03.dat},
  year         = {2020},
  howpublished = {\url{https://gitlab.com/mmoresco/CCcovariance/-/blob/master/data/HzTable_MM_BC03.dat}}
}

@article{Fred1996,
  title = {Nonconservative Lagrangian and Hamiltonian mechanics},
  author = {Riewe, Fred},
  journal = {Phys. Rev. E},
  volume = {53},
  issue = {2},
  pages = {1890--1899},
  numpages = {0},
  year = {1996},
  month = {Feb},
  publisher = {American Physical Society},
  doi = {10.1103/PhysRevE.53.1890}
}

@article{Lazo2014,
    author = {Lazo, Matheus J. and Krumreich, Cesar E.},
    title = {The action principle for dissipative systems},
    journal = {Journal of Mathematical Physics},
    volume = {55},
    number = {12},
    pages = {122902},
    year = {2014},
    month = {12},
    issn = {0022-2488},
    doi = {10.1063/1.4903991}
}

@article{ refId0,
	author = {{Planck Collaboration}},
	title = {Planck 2018 results - I. Overview and the cosmological legacy of Planck},
	DOI= "10.1051/0004-6361/201833880",
	journal = {A.A.},
	year = 2020,
	volume = 641,
	pages = "A1",
}

@article{CALDEIRA1983587,
title = {Path integral approach to quantum Brownian motion},
journal = {Physica A: Statistical Mechanics and its Applications},
volume = {121},
number = {3},
pages = {587-616},
year = {1983},
issn = {0378-4371},
doi = {https://doi.org/10.1016/0378-4371(83)90013-4},
url = {https://www.sciencedirect.com/science/article/pii/0378437183900134},
author = {A.O. Caldeira and A.J. Leggett}
}

@article{PhysRevD.111.066001,
  title = {Gravity from entropy},
  author = {Bianconi, Ginestra},
  journal = {Phys. Rev. D},
  volume = {111},
  issue = {6},
  pages = {066001},
  numpages = {17},
  year = {2025},
  month = {Mar},
  publisher = {American Physical Society},
  doi = {10.1103/PhysRevD.111.066001},
  url = {https://link.aps.org/doi/10.1103/PhysRevD.111.066001}
}

@article{Cieslinski_2010,
doi = {10.1088/1751-8113/43/17/175205},
url = {https://doi.org/10.1088/1751-8113/43/17/175205},
year = {2010},
month = {apr},
publisher = {},
volume = {43},
number = {17},
pages = {175205},
author = {Cieśliński, Jan L and Nikiciuk, Tomasz},
title = {A direct approach to the construction of standard and non-standard Lagrangians for dissipative-like dynamical systems with variable coefficients},
journal = {Journal of Physics A: Mathematical and Theoretical}
}

@article{Nojiri:2007bt,
    author = "Nojiri, Shin'ichi and Odintsov, Sergei D. and Tretyakov, Petr V.",
    editor = "Kenmoku, Masakatsu and Sasaki, Misao",
    title = "{From inflation to dark energy in the non-minimal modified gravity}",
    eprint = "0710.5232",
    archivePrefix = "arXiv",
    primaryClass = "hep-th",
    reportNumber = "YITP-07-73",
    doi = "10.1143/PTPS.172.81",
    journal = "Prog. Theor. Phys. Suppl.",
    volume = "172",
    pages = "81--89",
    year = "2008"
}

@article{Mohseni:2009ns,
    author = "Mohseni, Morteza",
    title = "{Non-geodesic motion in $f({\mathcal G})$ gravity with non-minimal coupling}",
    eprint = "0911.2754",
    archivePrefix = "arXiv",
    primaryClass = "hep-th",
    doi = "10.1016/j.physletb.2009.10.089",
    journal = "Phys. Lett. B",
    volume = "682",
    pages = "89--92",
    year = "2009"
}

@article{Harko:2018gxr,
    author = "Harko, Tiberiu and Koivisto, Tomi S. and Lobo, Francisco S. N. and Olmo, Gonzalo J. and Rubiera-Garcia, Diego",
    title = "{Coupling matter in modified $Q$ gravity}",
    eprint = "1806.10437",
    archivePrefix = "arXiv",
    primaryClass = "gr-qc",
    doi = "10.1103/PhysRevD.98.084043",
    journal = "Phys. Rev. D",
    volume = "98",
    number = "8",
    pages = "084043",
    year = "2018"
}

@article{Wu:2018idg,
    author = "Wu, Jimin and Li, Guangjie and Harko, Tiberiu and Liang, Shi-Dong",
    title = "{Palatini formulation of $f(R,T)$ gravity theory, and its cosmological implications}",
    eprint = "1805.07419",
    archivePrefix = "arXiv",
    primaryClass = "gr-qc",
    doi = "10.1140/epjc/s10052-018-5923-9",
    journal = "Eur. Phys. J. C",
    volume = "78",
    number = "5",
    pages = "430",
    year = "2018"
}

@article{Haghani:2021fpx,
    author = "Haghani, Zahra and Harko, Tiberiu",
    title = "{Generalizing the coupling between geometry and matter: $f\left( R,L_m,T\right) $ gravity}",
    eprint = "2106.10644",
    archivePrefix = "arXiv",
    primaryClass = "gr-qc",
    doi = "10.1140/epjc/s10052-021-09359-3",
    journal = "Eur. Phys. J. C",
    volume = "81",
    number = "7",
    pages = "615",
    year = "2021"
}

@article{deLimaJunior:2024icz,
    author = "de Lima J{\'u}nior, J. G. and Moraes, P. H. R. S. and Brito, E. and Fortunato, J. A. S.",
    title = "{The Palatini formalism of the $f(R,{\mathcal {L}}_{m},T)$ theory of gravity}",
    eprint = "2411.15615",
    archivePrefix = "arXiv",
    primaryClass = "gr-qc",
    doi = "10.1140/epjc/s10052-024-13731-4",
    journal = "Eur. Phys. J. C",
    volume = "85",
    number = "1",
    pages = "38",
    year = "2025"
}

@article{N1,
  author  = {Shen, Yuejun and Chen, Chutian and Ma, Haoran and 
             Saunders, Ashley P. and Heide, Christian and Liu, Fang and 
             Rotskoff, Grant M. and Shi, Jiaojian and Lindenberg, Aaron M.},
  title   = {Non-equilibrium entropy production and information dissipation in a non-Markovian quantum dot},
  journal = {Nature Physics},
  year    = {2026},
  doi     = {10.1038/s41567-026-03177-8},
  url     = {https://doi.org/10.1038/s41567-026-03177-8},
  issn    = {1745-2481}
}

@article{N2,
title = {Path integral approach to quantum Brownian motion},
journal = {Physica A: Statistical Mechanics and its Applications},
volume = {121},
number = {3},
pages = {587-616},
year = {1983},
issn = {0378-4371},
doi = {https://doi.org/10.1016/0378-4371(83)90013-4},
url = {https://www.sciencedirect.com/science/article/pii/0378437183900134},
author = {A.O. Caldeira and A.J. Leggett},
}

@article{N3,
  author  = {G{\"u}ttinger, Johannes and Noury, Adrien and Weber, Peter and 
             Eriksson, Axel Martin and Lagoin, Camille and Moser, Joel and 
             Eichler, Christopher and Wallraff, Andreas and 
             Isacsson, Andreas and Bachtold, Adrian},
  title   = {Energy-dependent path of dissipation in nanomechanical resonators},
  journal = {Nature Nanotechnology},
  volume  = {12},
  number  = {7},
  pages   = {631--636},
  year    = {2017},
  month   = {Jul},
  doi     = {10.1038/nnano.2017.86}
}

@inproceedings{N4,
  author      = {Maartens, Roy},
  title       = {Causal thermodynamics in relativity},
  booktitle   = {Proceedings of the Hanno Rund Workshop on Relativity and Thermodynamics},
  year        = {1996},
  eprint      = {astro-ph/9609119},
  archivePrefix = {arXiv},
  reportNumber  = {PU-RCG-96-14}
}

@article{N5,
doi = {10.1088/0034-4885/79/9/096901},
url = {https://doi.org/10.1088/0034-4885/79/9/096901},
year = {2016},
month = {aug},
publisher = {IOP Publishing},
volume = {79},
number = {9},
pages = {096901},
author = {Wang, B and Abdalla, E and Atrio-Barandela, F and Pavón, D},
title = {Dark matter and dark energy interactions: theoretical challenges, cosmological implications and observational signatures},
journal = {Reports on Progress in Physics},
}

@article{N6,
  doi = {10.22331/q-2023-10-16-1142},
  url = {https://doi.org/10.22331/q-2023-10-16-1142},
  title = {Any consistent coupling between classical gravity and quantum matter is fundamentally irreversible},
  author = {Galley, Thomas D. and Giacomini, Flaminia and Selby, John H.},
  journal = {{Quantum}},
  issn = {2521-327X},
  publisher = {{Verein zur F{\"{o}}rderung des Open Access Publizierens in den Quantenwissenschaften}},
  volume = {7},
  pages = {1142},
  month = oct,
  year = {2023}
}

@article{N7,
  title = {Irreversibility and gravitational radiation: A proof of Bondi's conjecture},
  author = {Herrera, L. and Di Prisco, A. and Ospino, J.},
  journal = {Phys. Rev. D},
  volume = {109},
  issue = {2},
  pages = {024005},
  numpages = {9},
  year = {2024},
  month = {Jan},
  publisher = {American Physical Society},
  doi = {10.1103/PhysRevD.109.024005},
  url = {https://link.aps.org/doi/10.1103/PhysRevD.109.024005}
}

@book{N8,
  author    = {Parker, Leonard and Toms, David},
  title     = {Quantum Field Theory in Curved Spacetime: Quantized Fields and Gravity},
  publisher = {Cambridge University Press},
  address   = {Cambridge},
  year      = {2009}
}

\end{document}